\documentclass[a4paper,11pt]{article}
\usepackage{jcappub}
\usepackage{amsmath}
\usepackage{graphicx}

\makeatletter
\gdef\@fpheader{}
\makeatother

\newlength{\fullw}
\newlength{\halfw}
\newlength{\twofigw}
\newlength{\onefigw}
\newcommand{\boldmathsymbol}[1]{{\ensuremath{\boldsymbol{#1}}}}
\newcommand{\sss}[1]{{\scriptscriptstyle{#1}}}
\newcommand{\vect}[1]{\boldmathsymbol{#1}}

\newcommand{\healpix}{\texttt{HEALPix}}
\newcommand{\camb}{\texttt{CAMB}}

\newcommand{\order}[1]{\mathcal{O}\!\left(#1\right)}
\newcommand{\dd}{\mathrm{d}}
\newcommand{\ud}{\dd}
\newcommand{\uc}{\mathrm{c}}

\newcommand{\K}{\mathrm{K}}

\newcommand{\GeV}{\mathrm{GeV}}
\newcommand{\Gpc}{\mathrm{Gpc}}
\newcommand{\muK}{\mu\mathrm{K}}

\newcommand{\uh}{\mathrm{h}}
\newcommand{\ueq}{\mathrm{eq}}
\newcommand{\ulss}{\mathrm{lss}}
\newcommand{\uini}{\mathrm{ini}}
\newcommand{\uside}{\mathrm{side}}
\newcommand{\uscal}{\mathrm{scal}}
\newcommand{\udelay}{\mathrm{thaw}}
\newcommand{\umat}{\mathrm{mat}}
\newcommand{\urad}{\mathrm{rad}}
\newcommand{\udip}{\mathrm{dip}}
\newcommand{\umon}{\mathrm{mon}}
\newcommand{\uobs}{\mathrm{obs}}
\newcommand{\LCDM}{\Lambda\mathrm{CDM}}
\newcommand{\uLCDM}{\sss{\LCDM}}

\newcommand{\calM}{\mathcal{M}}
\newcommand{\calD}{\mathcal{D}}

\newcommand{\U}{U}
\newcommand{\zlss}{z_{\ulss}}
\newcommand{\zrad}{z_{\urad}}
\newcommand{\zeq}{z_{\ueq}}
\newcommand{\unitn}{\vect{\hat{n}}}
\newcommand{\unitw}{\vect{\hat{\omega}}}
\newcommand{\Xp}{\acute{X}}
\newcommand{\Xd}{\dot{X}}
\newcommand{\Nside}{N_\uside}
\newcommand{\boxgpc}{L_\mathrm{sim}}
\newcommand{\corrini}{\ell_\uc}
\newcommand{\Ncorrini}{N_\uc}
\newcommand{\zero}{{\sss{0}}}

\newcommand{\dhini}{d_{\uh_\uini}}
\newcommand{\dhend}{d_{\uh_\zero}}
\newcommand{\OmegaM}{\Omega_\umat}
\newcommand{\OmegaR}{\Omega_\urad}
\newcommand{\mon}{\calM}
\newcommand{\dip}{\calD}

\title{Large scale CMB anomalies from thawing cosmic strings}

\author[a,b]{Christophe Ringeval,}
\author[c]{Daisuke Yamauchi,}
\author[c,d,e]{Jun'ichi Yokoyama}
\author[b]{and Fran\c{c}ois R. Bouchet}

\affiliation[a]{Centre for Cosmology, Particle Physics and Phenomenology,
  Institute of Mathematics and Physics, Louvain University, 2 Chemin
  du Cyclotron, 1348 Louvain-la-Neuve, Belgium}

\affiliation[b]{Institut d'Astrophysique de Paris, UMR
7095-CNRS, Universit\'e Pierre et Marie Curie, 98bis boulevard Arago,
75014 Paris, France}

\affiliation[c]{Research Center for the Early Universe (RESCEU), Graduate School
  of Science, The University of Tokyo, Tokyo 113-0033, Japan}

\affiliation[d]{Department of Physics, Graduate School of Science,
The University of Tokyo, Tokyo 113-0033, Japan.}

\affiliation[e]{Kavli Institute for the Physics and Mathematics of the
  Universe (Kavli IPMU), WPI, The University of Tokyo, Kashiwa, Chiba,
  277-8568, Japan}

\emailAdd{christophe.ringeval@uclouvain.be}
\emailAdd{yamauchi@resceu.s.u-tokyo.ac.jp}
\emailAdd{yokoyama@resceu.s.u-tokyo.ac.jp}
\emailAdd{bouchet@iap.fr}

\date{today}

\begin{document}

\abstract{Cosmic strings formed during inflation are expected to be
  either diluted over super-Hubble distances, i.e., invisible today,
  or to have crossed our past light cone very recently. We discuss the
  latter situation in which a few strings imprint their signature in
  the Cosmic Microwave Background (CMB) Anisotropies after
  recombination. Being almost frozen in the Hubble flow, these strings
  are quasi static and evade almost all of the previously derived
  constraints on their tension while being able to source large scale
  anisotropies in the CMB sky. Using a local variance estimator on
  thousand of numerically simulated Nambu-Goto all sky maps, we
  compute the expected signal and show that it can mimic a dipole
  modulation at large angular scales while being negligible at small
  angles. Interestingly, such a scenario generically produces one cold
  spot from the thawing of a cosmic string loop. Mixed with
  anisotropies of inflationary origin, we find that a few strings of
  tension $G\U = \order{1} \times 10^{-6}$ match the amplitude of the
  dipole modulation reported in the Planck satellite measurements and
  could be at the origin of other large scale anomalies.}


\maketitle

\section{Introduction}
\label{sec:intro}

Although cosmic strings are a natural outcome of the symmetry breaking
mechanism in the early Universe~\cite{Kirzhnits:1972, Kibble:1976}, or
may be more fundamental objects from String
Theory~\cite{Witten:1985fp, Dvali:1998pa}, they have escaped all
dedicated searches in astronomical and cosmological observables to
date (see Refs.~\cite{Hindmarsh:1994re, Durrer:2002,
  Polchinski:2004ia, Davis:2008dj, Copeland:2009ga,
  Sakellariadou:2009ev, Ringeval:2010ca} for reviews). Being active
and non-Gaussian sources of CMB anisotropies~\cite{Contaldi:1998mx,
  Hindmarsh:2009qk, Hindmarsh:2009es, Regan:2015cfa}, current
measurements by the Planck satellite allow strings to contribute no
more than a few percent to the overall power spectrum and
bispectrum. This translates into an $95\%$ confidence upper bound on
the allowed string tension: $G\U \le 3.2 \times 10^{-7}$ for Abelian
Higgs string and $G\U < 1.5 \times 10^{-7}$ for Nambu-Goto
strings~\cite{Urrestilla:2011gr, Ade:2013xla, Lizarraga:2014xza,
  Lazanu:2014xxa, Lazanu:2014eya}. It is important to notice that
these results rely on the assumption that a cosmic string network
evolves according to the so-called scaling regime. Indeed, in an
expanding Universe, it has been shown that after a few e-folds of
decelerated expansion the distribution of cosmic strings reaches an
attractor in which the network's statistical properties become an
universal function of the Hubble radius only~\cite{Albrecht:1989,
  Bennett:1989, Allen:1990, Bennett:1990}. Typically, the scaling
regime for Nambu-Goto strings corresponds, at any times, to ten long
strings crossing the observable Universe complemented with a power law
distribution of smaller loops~\cite{Vanchurin:2005yb, Ringeval:2005kr,
  Lorenz:2010sm, Blanco-Pillado:2013qja, Peter:2013jj}. Assuming
``scaling'' is usually well motivated for observational purposes. If
strings are formed around the Grand Unified Theory (GUT) energy
scales, there is indeed plenty of time for any initial configuration
to relax towards the attractor.

Within our current understanding of the early Universe, there are
various explanations for strings to remain elusive in the cosmological
measurements. The simplest is that their tension is relatively small
compared to GUT scale, typically lower than $(10^{15}\,\GeV)^2$, and they
remain undetectable for the moment. Another explanation is that, as
for monopoles, cosmic strings may have been diluted away by cosmic
inflation, provided they have been formed before inflation. Or, they
may have never been formed at all. Both explanations would however
require some fine tuning within the symmetry breaking schemes
at work in the cooling of the fundamental
interactions~\cite{Jeannerot:2003qv, Rocher:2004my}.

Here we discuss another proposal which is that cosmic strings may have
been formed during inflation~\cite{Lazarides:1984pq, Shafi:1984tt,
  Vishniac:1986sk, Kofman:1986wm, Yokoyama:1988zza, Yokoyama:1989pa,
  Nagasawa:1991zr, Basu:1993rf, Freese:1995vp, Kamada:2012ag,
  Zhang:2015bga}. As discussed in Ref.~\cite{Kamada:2014qta}, the
network statistical properties are no longer universal as they depend
on when and how the strings have been formed during inflation. If they
have been formed more than, say, $60$ e-folds before the end of
inflation (the precise number is inflation- and reheating-dependent),
strings would be diluted enough not to be observable at all. If, on
the contrary, they have been formed closer to the end of inflation,
the typical correlation length of the string network would remain
super-Hubble for most of the radiation and matter eras of the
Universe, but will eventually enter the Hubble radius at some
point. Following Ref.~\cite{Kamada:2014qta}, we refer to this scenario
as delayed scaling. As shown in this reference, delayed scaling
strings induce less CMB anisotropies than a scaling network because
less strings intercept our past light cone. In particular, the
reduction in power preferentially occurs on the smallest length scales
(large multipoles). As a result, these scenarios naturally imprint the
CMB anisotropies on large scales only and one may wonder whether they
could be a viable explanation of the large scales anomalies reported
in the Wilkinson Microwave Anisotropies Probe and Planck
measurements~\cite{Ade:2013nlj, Notari:2013iva, Rassat:2014yna,
  Quartin:2014yaa, Ade:2015sjc}.

In the following, we focus on a delayed scaling scenario in which the
strings cross our past light cone after recombination. In order to
derive their observable imprints, we use numerical simulations of
Nambu-Goto strings in which we propagate photons to ray-trace the CMB
sky using a method similar to the one of Ref.~\cite{Ringeval:2012tk}
and described in section~\ref{sec:simus}. In particular, we find that
starting with a Vachaspati-Vilenkin string
network~\cite{Vachaspati:1984} having a super-Hubble correlation
length at last scattering, $\xi \simeq 32 \eta_{\ulss}$ (in comoving
coordinates, $\eta$ being the conformal time) our past light cone
would contain from zero to two strings which remain almost static till
today. The resulting CMB angular power spectrum ends up being at least
two orders of magnitude lower than the one produced by a scaling
network such that the scenario ends up being constrained only by
direct searches of rare Gott-Kaiser-Stebbins~\cite{Gott:1984ef,
  Kaiser:1984iv} (GKS) temperature discontinuities~\cite{Bouchet:1988,
  Fraisse:2007nu, Danos:2008fq, Danos:2009vv, Jeong:2010ft}. As
discussed below, this typically allows string tension up to $G\U =
\order{1} \times 10^{-6}$ to be compatible with current
constraints. In section~\ref{sec:mod}, we quantify to which extent
these strings can generate a power asymmetry in the CMB sky by using a
local variance estimator introduced in
Ref.~\cite{Akrami:2014eta}. From a thousand random realizations of the
string network with $\xi \simeq 32 \eta_{\ulss}$, we find that the
induced CMB patterns can, among other effects, generically generate a
large scale dipole modulation in the local variance. We also show that
the amplitude of the modulation is very peculiar to this
scenario and we discuss how it is modified by changing the initial
value of $\xi$ in section~\ref{sec:thawing}. In particular, for an
initial correlation length $\xi \simeq 16 \eta_{\ulss}$, our light
cone can contain a thawing cosmic string loop which genuinely
generates a cold spot in the CMB sky. Finally, it is found that string
tensions around $G\U =\order{1} \times 10^{-6}$ are found to mimic the
currently observed dipole modulation in the Planck
data~\cite{Ade:2013nlj, Akrami:2014eta}.

\section{All sky CMB maps}
\label{sec:simus}

\subsection{Method}
\label{sec:method}

In order to derive the CMB temperature anisotropies generated by
cosmic strings formed during inflation, we have performed numerical
simulations of Nambu-Goto strings evolution in
Friedmann-Lema\^{\i}tre-Robertson-Walker (FLRW) spacetimes. The code
used is based on an improved version of the Bennett and Bouchet
Nambu-Goto cosmic string code~\cite{Bennett:1990}. For our purpose,
the initial conditions have been set using the Vachaspati-Vilenkin
algorithm~\cite{Vachaspati:1984} which attributes a random phase
(correlated over a distance $\xi$) to an hypothetical $U(1)$ Higgs
field on a three-dimensional cubic lattice. Depending on the phase
topology around each corners of the cubic grid, the algorithm
determines if a string actually passes, or not, through each of the
cubic faces. Because the strings are assumed to be formed during
inflation, we have set vanishing initial velocity to each of the
strings and loops thus generated. The fundamental difference compared
to the more ordinary situation in which one evolves a string network
in scaling is that we now set the initial correlation length of Higgs
phases to be much larger than the Hubble radius. In order to simulate
the actual physical configuration, it would be necessary to set the
correlation length during inflation and solve the Nambu-Goto dynamics
for about $60$ e-folds of inflation and as much e-folds of decelerated
expansion during the subsequent reheating, radiation and matter
eras. Of course, this is not doable numerically but it is not a
limitation as the Nambu-Goto dynamics becomes trivial in the
super-Hubble limit $\xi/\eta\rightarrow\infty$ ($\eta$ being the
conformal time). Indeed, strings remain frozen due to Hubble
damping~\cite{Avelino:2007iq} and the overall configuration of the
network is straightforwardly
redshifted~\cite{Kamada:2014qta}. Therefore, we have chosen to start
the simulation in the matter era at a redshift close to the last
scattering surface $z=\zlss=1089$. Our initial configuration is
described by $\left. \xi/\eta \right|_{\ulss}$, which is an observable
model parameter for the delayed scaling scenario, and gives the
correlation length of the network at the time of last scattering. From
the above discussion, such an approach is valid only if $\xi >
\eta_\ulss$ to prevent any significant difference in the network
evolution before the last scattering surface with respect to simple
redshifting. The resulting CMB anisotropies are completely sourced by
the strings after recombination and can be computed from the
integrated Sachs-Wolfe effect associated with the Nambu-Goto stress
tensor of the whole network. In the transverse temporal gauge, up to a
dipole term, the relative photon temperature shifts in the direction
$\unitn$ are given by~\cite{1992ApJ395L55V, Hindmarsh:1993pu,
  Stebbins:1995}.
\begin{equation}
\label{eq:isw}
\dfrac{\Delta T}{T}(\unitn) = - 4 G \U \int_{\vect{X}\,\cap\,\vect{x}_\gamma} 
\vect{u} \cdot \dfrac{X \unitn - \vect{X}}{\left(X \unitn -
    \vect{X} \right)^2} \,\epsilon\, \ud \sigma,
\end{equation}
where the integral is over all string position vectors
$\vect{X}=\{X^i\}$ crossing our past line cone. Here $\ud l=\epsilon
\ud \sigma$ is the invariant string length element and $\vect{u}$ the
sourcing vector of the Gott-Kaiser-Stebbins effect (in the temporal
gauge):
\begin{equation}
\epsilon \equiv \sqrt{\dfrac{\vect{\Xp}^2}{1-\vect{\Xd}^2}}\,, \qquad
\vect{u} = \vect{\Xd} - \dfrac{(\unitn \cdot \vect{\Xp}) \cdot
  \vect{\Xp}}{1 + \unitn \cdot \vect{\Xd}}\,.
\label{eq:dtot}
\end{equation}
Both $\vect{\Xd}=\partial \vect{X}/\partial \eta$ and
$\vect{\Xp}=\partial \vect{X}/\partial \sigma$ are extracted from the
numerical simulations while the set of string position vectors
$\vect{X}$ intercepting our past light cone is determined by
propagating photons within the simulation box (see
Ref.~\cite{Ringeval:2012tk} for more details).

\subsection{Nambu-Goto numerical simulations}

\begin{figure}
\begin{center}
\includegraphics[width=\twofigw]{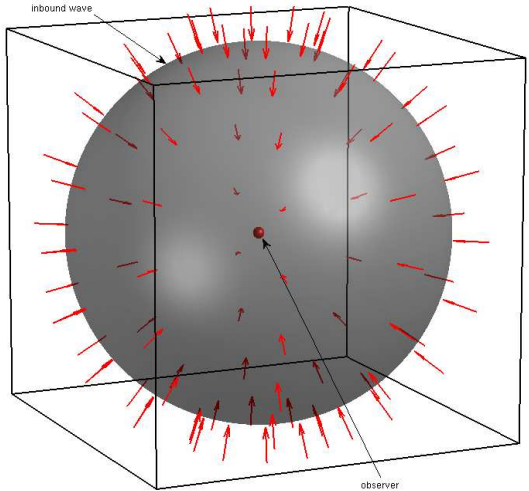}
\caption{Sketch of the numerical simulation comoving box. An inbound
  spherical light-like wave is propagated towards the centre during
  the run to determine the string position vectors $\vect{X}$
  intercepting our past light cone.}
\label{fig:box}
\end{center}
\end{figure}

In practice, our runs are performed in a unit comoving box in which we
are free to set two dimensionless numerical parameters (normalized by
the comoving size of the numerical box): the initial conformal time
(or conformal horizon), $\dhini$ and the initial correlation length
$\corrini$. In order to implement the delayed scaling string scenario
described before, one must have: $\corrini/\dhini = \xi/\eta_\ulss >
1$. Therefore, compared to the method described in
Ref.~\cite{Ringeval:2012tk}, which was dedicated to a scaling network
having initially $\left. \corrini/\dhini\right|_{\uscal} < 1$, our
simulation box contains only a very few number of strings. This
renders possible to actually simulate at once the whole observable
universe from the last scattering surface to today. At the same time
we evolve the string network according the the Nambu-Goto equations in
FLRW background, an inbound spherical wave of photons is propagated
through the simulation. By recording all string segments intercepting
this wave, one can reconstruct the past light cone of a virtual
observer located in the centre of the sphere (see
figure~\ref{fig:box}). Performing the integral of Eq.~\eqref{eq:isw}
over all directions $\unitn$ gives the string induced CMB sky the
observer would see. However, this also limits the time during which
the simulation can be run: an inbound light-like wave initially
inscribed in the unit comoving box collapses onto the observer at the
numerical conformal time $\dhend = \dhini+0.5$. Since we want to
compute the network evolution over a redshift range going up to the
last scattering surface, $\zlss=1089$, this imposes $\dhend/\dhini
\simeq \sqrt{1+\zlss}$. Solving for $\dhini$ one gets
\begin{equation}
\dhini \simeq \dfrac{1}{2 \left(\sqrt{1+\zlss} - 1 \right)} \simeq
1.56\times 10^{-2} \,,
\end{equation}
a value that determines the real comoving size of the numerical
box\footnote{This is an analytic approximation assuming no
  cosmological constant and $\OmegaR \ll \OmegaM$. The actual value
  used to determine the angular size of string
  generated CMB patterns is computed exactly within the $\Lambda$CDM
  model.} (see Refs.~\cite{Fraisse:2007nu, Ringeval:2012tk})
\begin{equation}
\boxgpc \simeq \dfrac{2}{\dhini H_0} \dfrac{1 -
  \sqrt{\OmegaR/\OmegaM} \sqrt{1+z_\ulss}}
{\sqrt{\OmegaM} \sqrt{1 + z_\ulss}} \,\Gpc \simeq 20.6\,\Gpc\,.
\label{eq:boxgpc}
\end{equation}
$H_0$ stands for the Hubble parameter today and $\OmegaR$ and
$\OmegaM$ are the density parameters today of radiation and matter.

The only remaining free parameter is $\corrini$. Because the initial
conditions are set according to the Vachaspati-Vilenkin algorithm,
namely over a comoving grid of the simulation box, $\corrini =
1/\Ncorrini$, where $\Ncorrini > 1$ is an integer. Our simulations can
therefore be used to evolve an initial configuration having
\begin{equation}
\left.\dfrac{\xi}{\eta} \right|_{\ulss} = \dfrac{1}{\Ncorrini \dhini}
\simeq \dfrac{64}{\Ncorrini}\,,
\end{equation}
the maximal value being obtained for $\Ncorrini=2$,
i.e., $\left. \xi/\eta \right|_{\ulss} \simeq 32$. One may compare this
number to the ones used in Ref.~\cite{Kamada:2014qta} in which a
network of super-Hubble strings was considered at an initial redshift
of $\zrad=2.3\times 10^7$. With a mixture of matter and radiation, one
gets
\begin{equation}
\left. \dfrac{\xi}{\eta} \right|_{\urad} \simeq
\dfrac{\sqrt{1+\dfrac{1+\zeq}{1+\zlss}}-1}{\sqrt{1+\dfrac{1+\zeq}{1+\zrad}}-1}
\left. \dfrac{\xi}{\eta} \, \right|_{\ulss} \simeq 5 \times 10^5.
\end{equation}
Our study is therefore probing a regime for which strings are formed
relatively earlier during inflation than the cases discussed in
Ref.~\cite{Kamada:2014qta}. As such they spend most of their evolution
in a frozen super-Hubble configuration during which the network length
scale gradually approaches the Hubble scale. At the time these two
scales are matching, strings start to decouple from the Hubble flow
and move under their tension. We will be referring to this regime as
``thawing'' because only after the correlation length falls well below
the Hubble radius, the system will enter the delayed scaling regime
studied in Ref.~\cite{Kamada:2014qta}, and ultimately the usual
scaling regime. In our treatment, only the thawing strings can be seen
in the sky and contribute to the CMB anisotropies. CMB signatures on
small scales disappear almost completely for such large values of the
correlation length but one may still expect some CMB distortions to be
present on the largest length scales. This is quantified in the
following sections.

\subsection{Temperature maps}

\begin{figure}
\begin{center}
\includegraphics[angle=90,width=\twofigw]{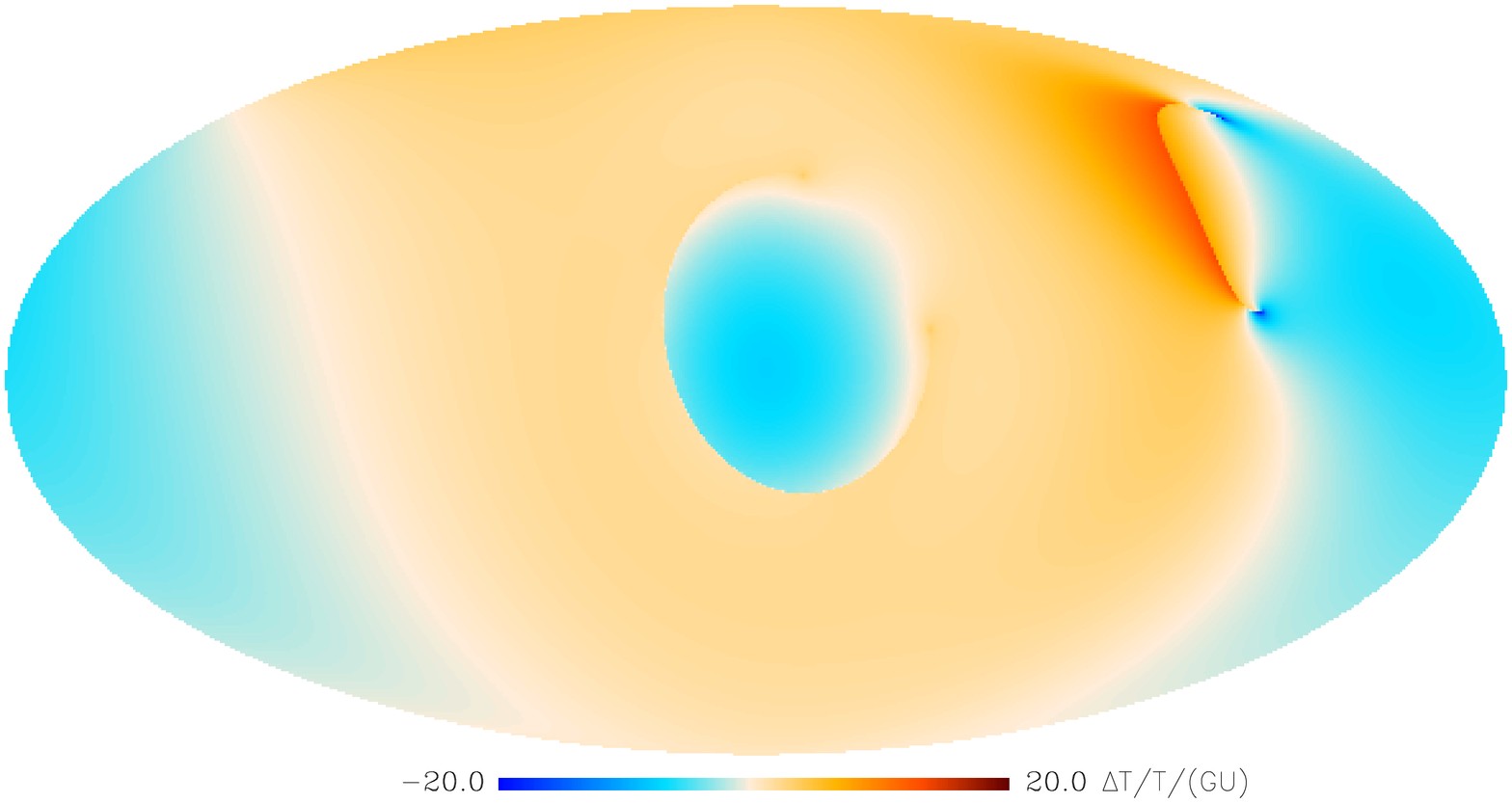}
\includegraphics[angle=90,width=\twofigw]{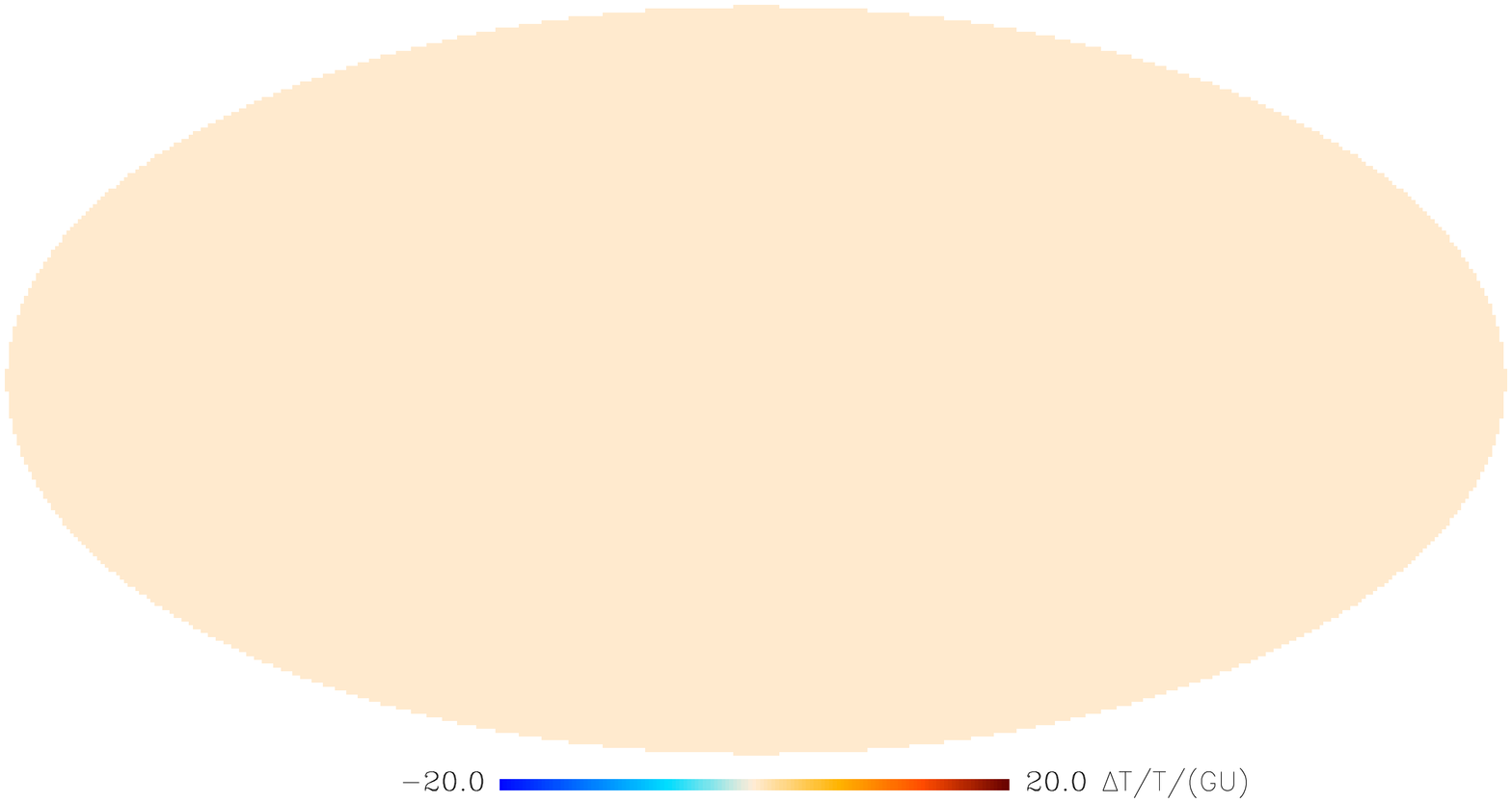}
\includegraphics[angle=90,width=\twofigw]{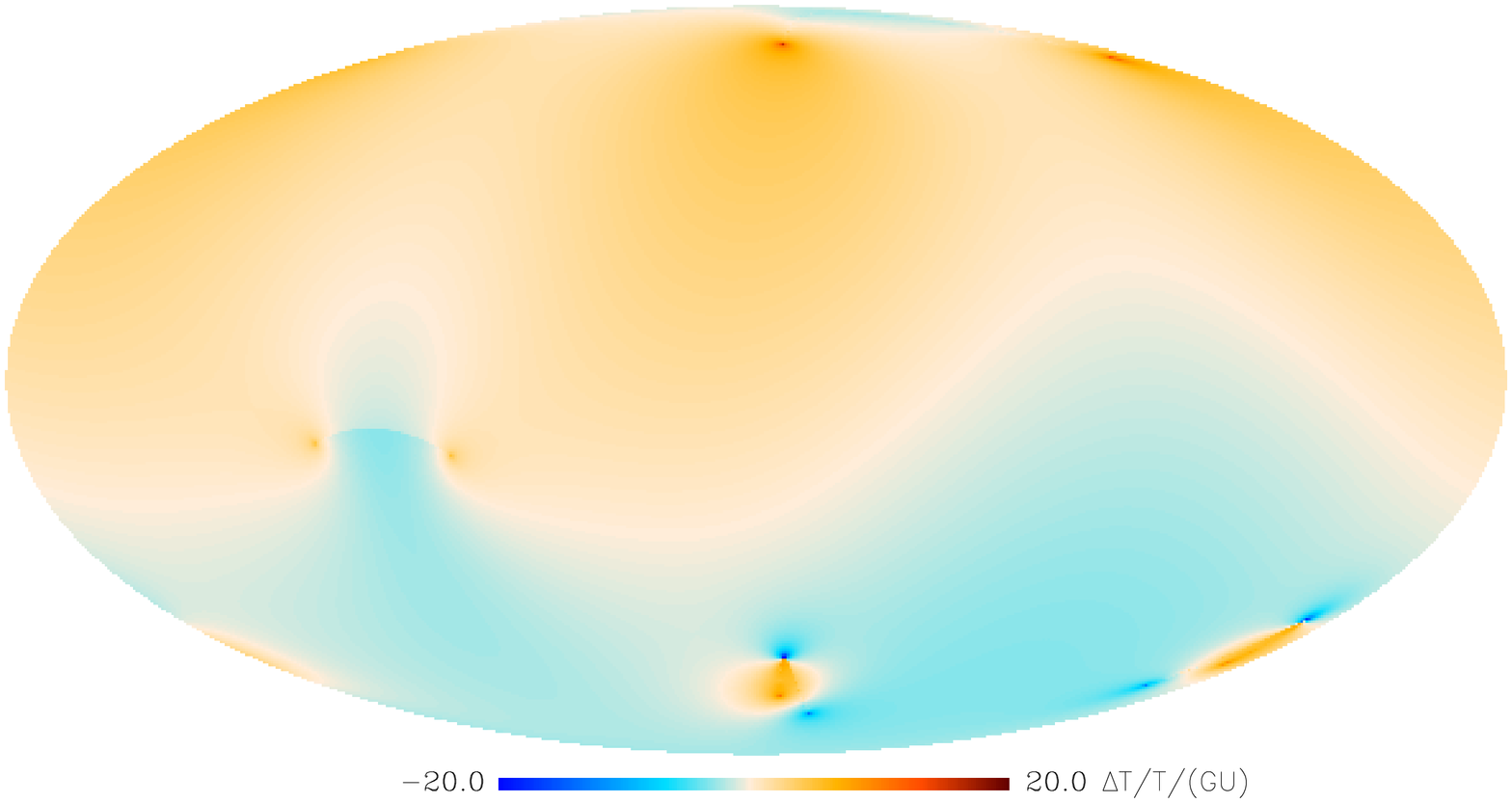}
\includegraphics[angle=90,width=\twofigw]{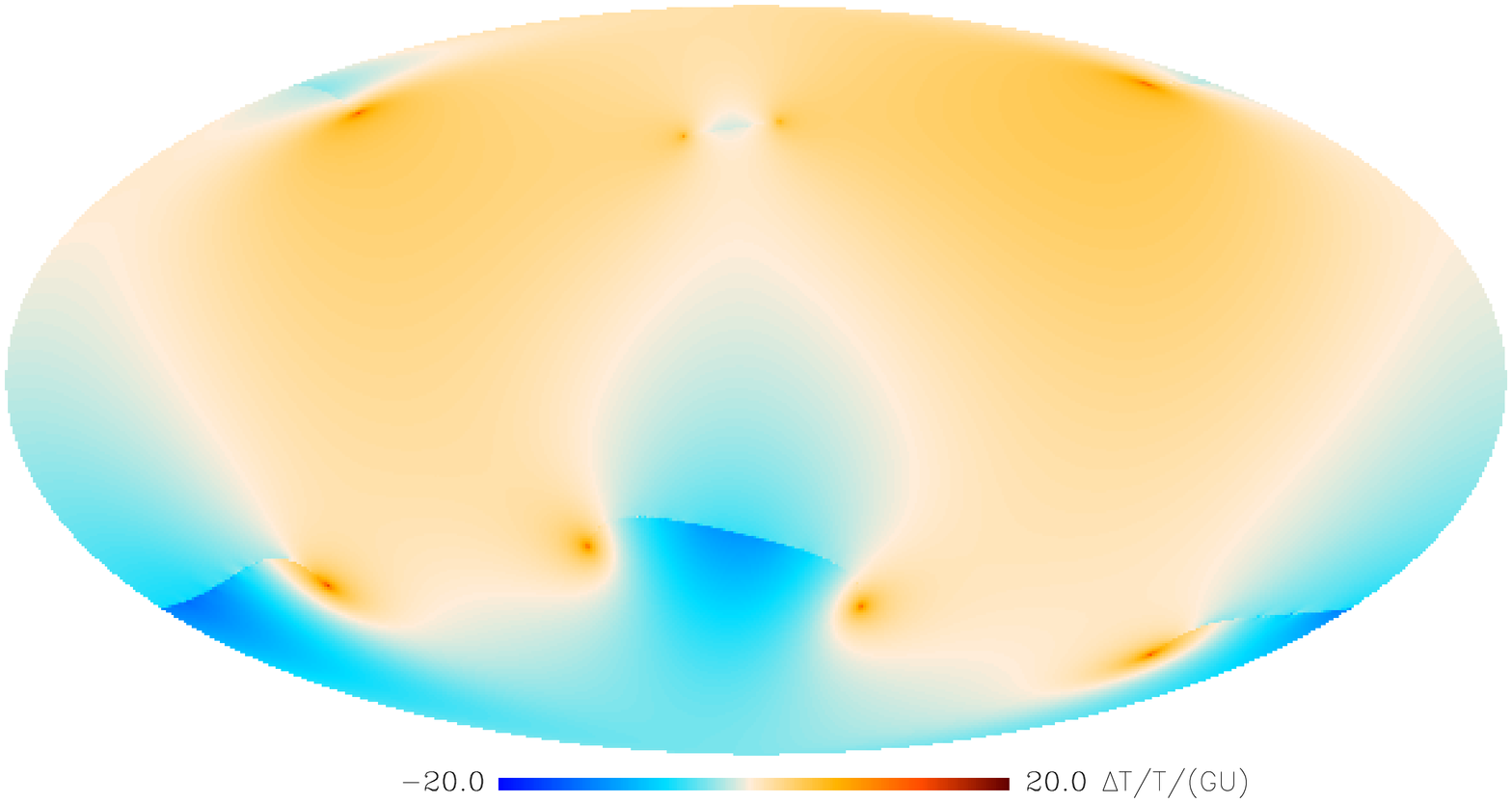}
\includegraphics[angle=90,width=\twofigw]{stg_n2_0001}
\includegraphics[angle=90,width=\twofigw]{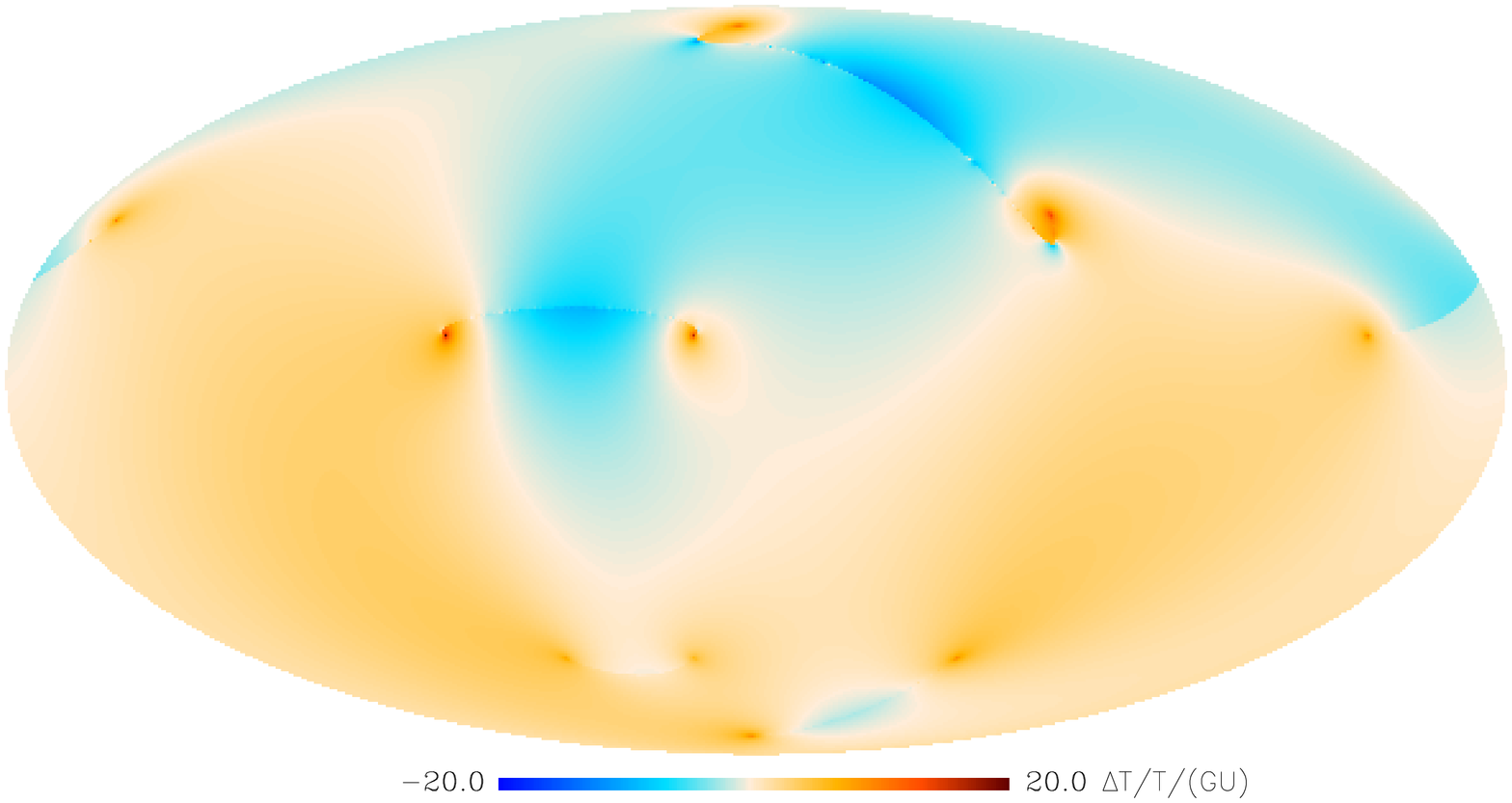}
\caption{Six realizations of the CMB sky temperature generated by
  thawing scaling strings having an initial correlation length $\xi =
  32 \eta_\ulss$ at the last scattering surface ($\Ncorrini=2$). In
  the second and fourth map (from top left to bottom right), the
  signal vanishes because no string crossed our past light cone. In
  the other panels, the temperature patterns are generated by a few
  loops (see first map) entering the Hubble radius and starting to
  evolve according to the Nambu-Goto dynamics. Sometimes, part of it
  intercept the last scattering surface which is why some temperature
  patterns seem to be attached on it. We have generated a thousand of
  such maps to compute their statistical properties.}
\label{fig:stgmaps2}
\end{center}
\end{figure}

The CMB temperature anisotropies generated by the strings are obtained
by the method described in section~\ref{sec:method}. In order to match
the Planck angular resolution, the sky has been pixelized over
more than $50\,000\,000$ directions ($\Nside = 2048$) using the {\healpix}
library~\cite{Gorski:2004by}. Moreover, in order to discuss the
statistical properties and cosmic variance effects, we have generated
$1024$ realizations of these maps by running numerical
simulations of Nambu-Goto strings starting from independent random
realizations of the initial conditions, and random positions of the
observer within the simulation volume. This is particularly relevant
as the statistics of the CMB anisotropies sourced by cosmic strings is
genuinely non-Gaussian~\cite{Ringeval:2010ca}. As discussed below, a
few strings affecting the largest CMB angular scales can significantly
boost cosmic variance effects thereby changing the probability of rare
events compared to what one may expect from purely Gaussian
anisotropies.

In figure~\ref{fig:stgmaps2}, we have represented six different
realizations of the CMB temperature anisotropies, all having an
initial value of $\Ncorrini=2$, i.e., $\xi/\eta|_\ulss \simeq 32$. As
can be seen on the upper right panel, in some realizations, no string
actually crosses our past light cone whereas in other situations one
or two strings are visible, either completely (as in the top left
panel) or some part of it (middle left).

\subsection{Angular power spectrum and constraints}
\label{sec:directobs}

\begin{figure}
\begin{center}
\includegraphics[width=\onefigw]{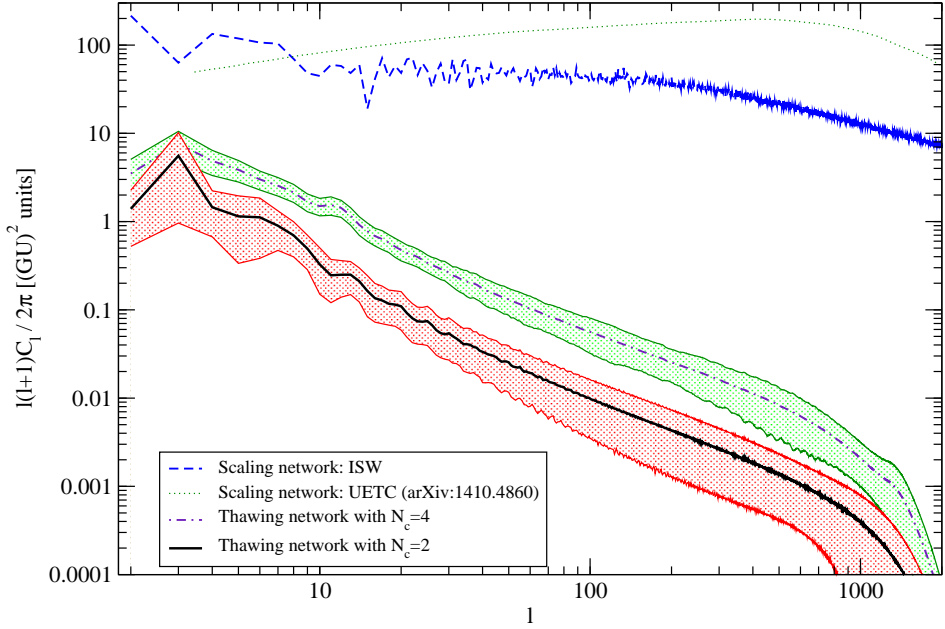}
\caption{Mean and standard deviation of the CMB temperature angular
  power spectrum for the delayed scaling string network in the thawing
  regime with $\Ncorrini=2$ and $\Ncorrini=4$ (convolved with a
  Gaussian beam of $10'$). For comparison, we have represented the
  total angular power spectrum for a Nambu-Goto network in scaling
  obtained using the UETC method (extracted from
  Ref.~\cite{Lazanu:2014xxa}) as well as the ISW contribution obtained
  from ray tracing~\cite{Ringeval:2012tk}. Thawing strings with
  $\Ncorrini=2$ have an angular power spectrum at least two orders of
  magnitude smaller than scaling strings.}
\label{fig:stgcls24}
\end{center}
\end{figure}

As above-mentioned, for $\Ncorrini=2$, the strings have barely the
time to start evolving after entering the Hubble radius and are
quasi-static. Therefore, the induced CMB temperature anisotropies are
orders of magnitude smaller than the ones generated by a scaling
network~\cite{Ade:2013xla}. In fact, for quasi-static strings, the induced
temperature anisotropies are essentially sourced by the curvature term
of $\vect{u}$ in equation~\eqref{eq:dtot}. In
figure~\ref{fig:stgcls24}, we have represented the mean angular power
spectrum, and its standard deviation, obtained over the $1024$
generated CMB maps. This figure also shows the angular power spectrum
obtained from Nambu-Goto strings in scaling, as obtained from the
Unequal Time Correlator (UETC) method in
Ref.~\cite{Lazanu:2014xxa}. We have also represented the Integrated
Sachs-Wolfe (ISW) contribution of scaling strings as obtained from the
ray-tracing method of Ref.~\cite{Ringeval:2012tk}. The ratio between
the spectra of thawing strings and scaling strings at $\ell=10$ is of
the order $C_{10}^{\udelay}/C_{10}^{\uscal} \simeq 4\times 10^{-3}$
while around the maximum power for scaling strings, at $\ell=400$,
$C_{400}^{\udelay}/C_{400}^{\uscal} \simeq 2.5 \times 10^{-5}$. The
two-sigma upper bound on the scaling string tension obtained with the
Planck data is of the order $G\U \simeq 1.5 \times
10^{-7}$~\cite{Ade:2013xla, Lazanu:2014xxa}, mostly coming from the
angular scales around $\ell=400$ where cosmic variance effects are
minimal. From $C_\ell \propto (G\U)^2$, one deduces that delayed
scaling strings in the thawing regime with $G\U=\order{1}\times
10^{-6}$ remain invisible in the angular power spectrum. One may
compare the resultant power spectra to those obtained in
Ref.~\cite{Kamada:2014qta}, in which the evolution of the delayed
scaling string network was assumed to follow the velocity-dependent
one-scale model. The amplitude of the power spectrum turns out to be
several tens of times larger than the one obtained here in our
simulations. This discrepancy is mainly due to the different regimes
probed. The analytic model cannot be extrapolated to the thawing
regime because strings have not yet interacted enough to correlate
their velocity and lengths scales. Moreover, when strings are almost
static, the GKS temperature
anisotropy~\cite{Gott:1984ef,Kaiser:1984iv} becomes dominated by
curvature and point-like sources such as kinks, cusps, and string
end-points on the last scattering surface. These contributions are
responsible for a slope at large multipoles slightly steeper than the
one predicted by the usual line discontinuities (see
figure~\ref{fig:stgcls24}).

\begin{figure}
\begin{center}
\includegraphics[angle=90,width=\twofigw]{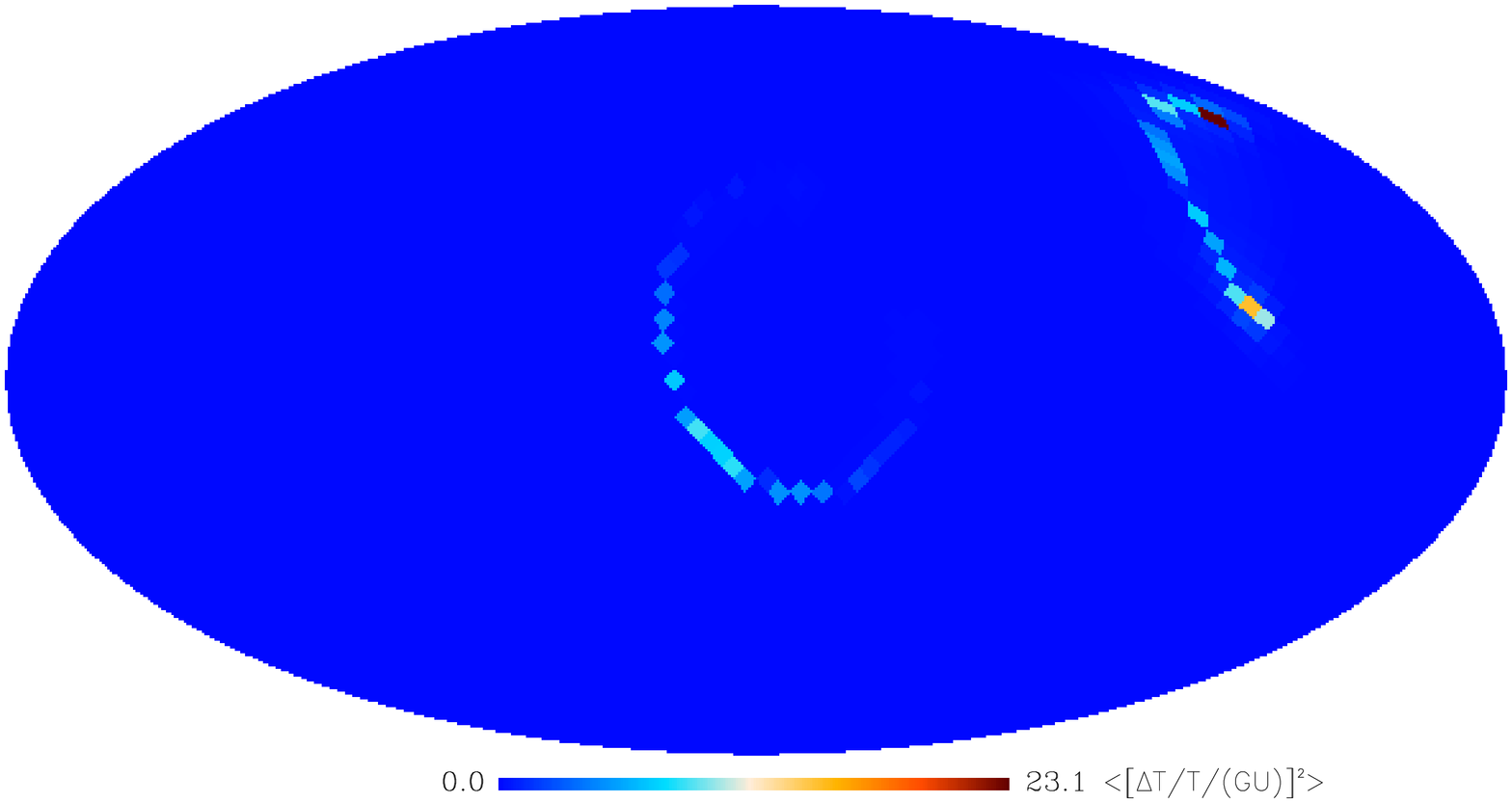}
\includegraphics[angle=90,width=\twofigw]{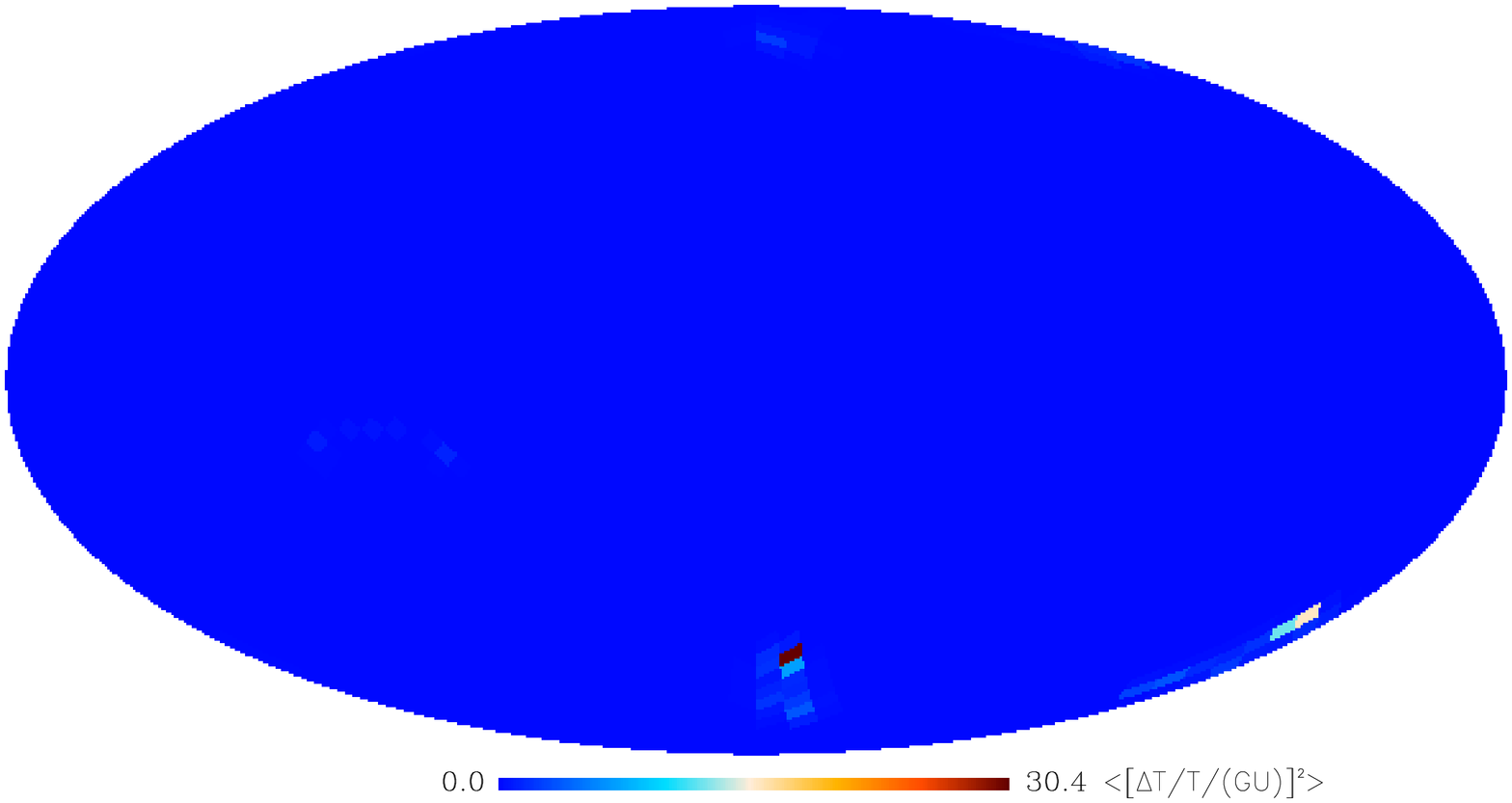}
\includegraphics[angle=90,width=\twofigw]{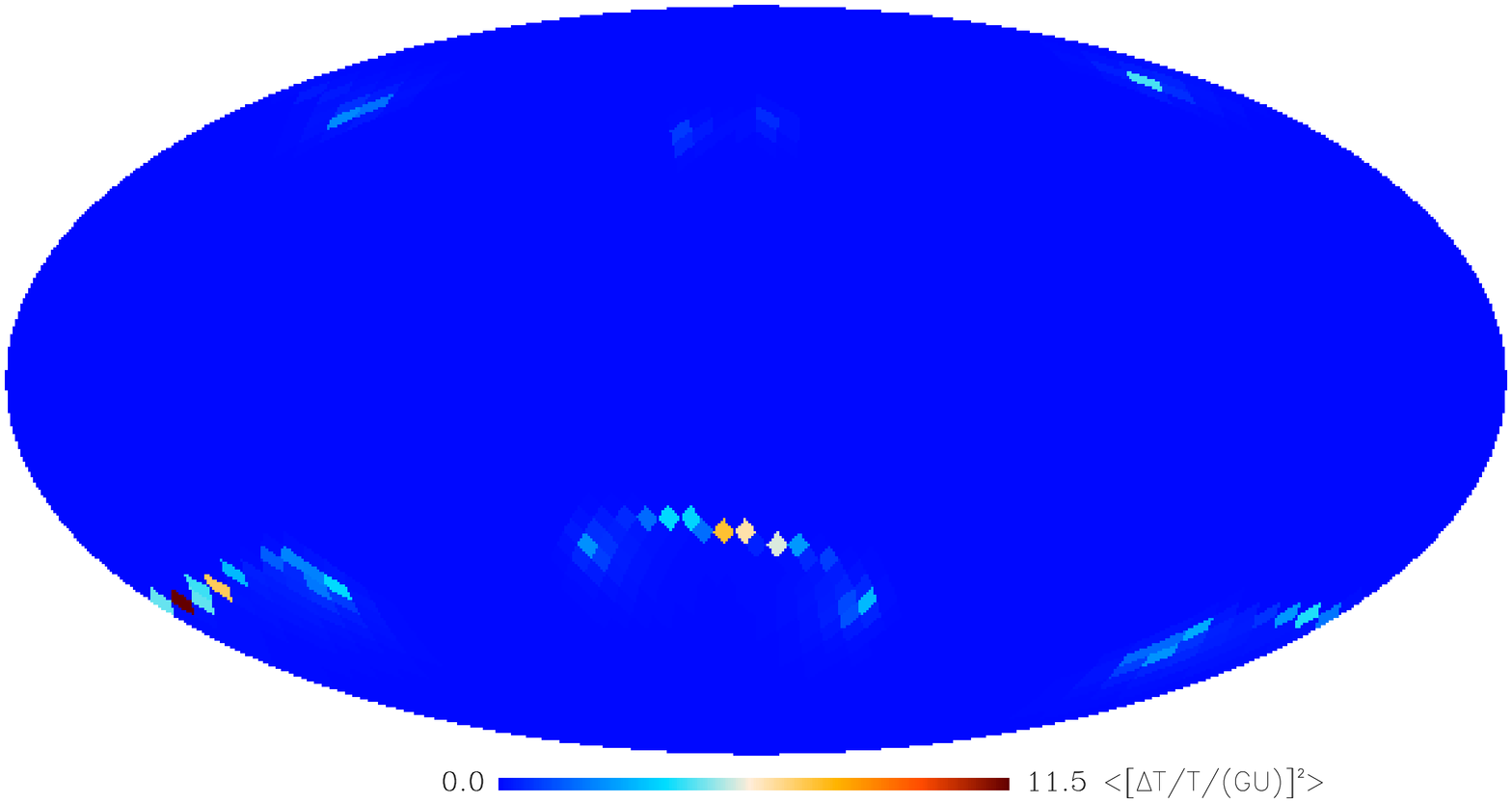}
\includegraphics[angle=90,width=\twofigw]{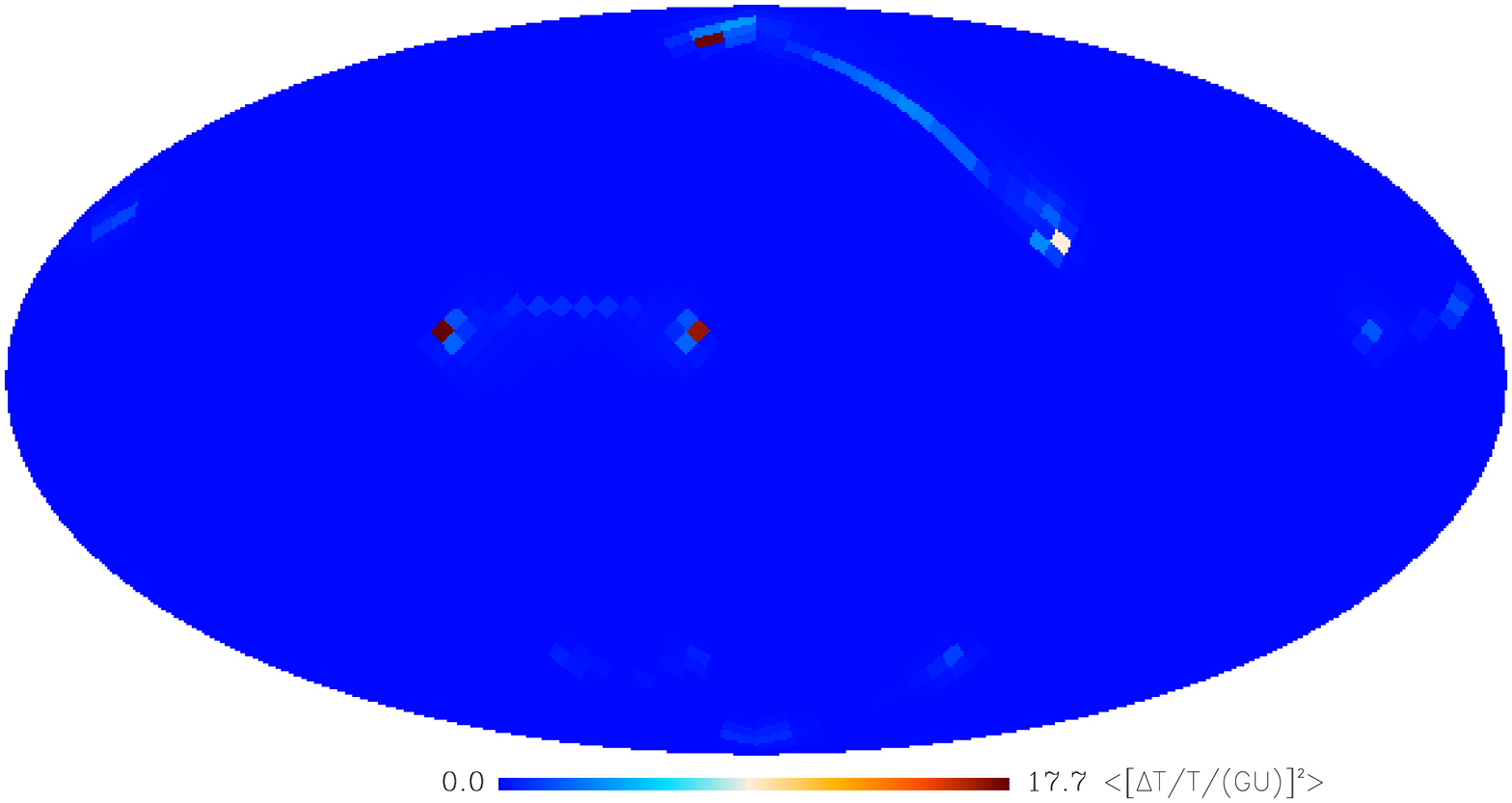}
\caption{Local variance maps of $\langle \left[\Delta T/(T
    G\U)\right]^2\rangle$ over $1.7^\circ$ circles associated with the
  four non-vanishing temperature maps of
  figure~\ref{fig:stgmaps2}. The maximal value of the standard
  deviation $\sigma$ is reached along one or two directions only and
  equals $4.8$, $5.5$, $3.4$ and $4.2$ for the four maps,
  respectively.}
\label{fig:stgvar2}
\end{center}
\end{figure}

However, as can be seen in figure~\ref{fig:stgmaps2}, for
$G\U=\order{1}\times 10^{-6}$, the CMB sky would contain one or two
string-induced temperature discontinuities having an amplitude of tens
of $\mu \K$ that could be detectable with direct
searches~\cite{Jeong:2007, Jeong:2010ft}. More precisely,
Ref.~\cite{Jeong:2010ft} reports that string induced discontinuities
of amplitude $\Delta = 117\,\muK$ over patches of $1.7^\circ$ would be
detectable in the Planck data at $95\,\%$ of confidence. Notice that
this result is not directly applicable to our situation as the
searches performed in Refs.~\cite{Jeong:2007, Jeong:2010ft} are made
for scaling-like strings, namely it is assumed that there are a few
strings moving at relativistic velocities in every $1.7^\circ$ patches
in the sky. Here, the strings are almost static and there are only one
or two temperature steps over the whole sky.

In order to estimate what is the amplitude of the temperature steps
produced over a given angular scale in the thawing regime, we have
represented in figure~\ref{fig:stgvar2} four local variance maps
computed over patches of angular opening $1.7^\circ$. Each of them is
associated with the four non-vanishing CMB temperature maps of
figure~\ref{fig:stgmaps2}, i.e., corresponds to a network having
$\Ncorrini=2$. Practically, we have discretized the sky using the
{\healpix} pixelization scheme with $\Nside=16$, which corresponds to
$3072$ directions. In each of these directions, the variance
$\sigma^2=\langle \left[\Delta T/(T G \U) \right]^2 \rangle$ is
computed over a circle having an angular opening of $1.7^\circ$. If no
temperature step is present in that circle, $\sigma^2$ remains very
small whereas if a string crosses the patch, $\sigma^2$ is
proportional to $\Delta^2$. The maximal variance being obtained for a
string cutting the circle in two equal parts~\cite{Jeong:2010ft}, one
has
\begin{equation}
\Delta = \max \left(2 \sigma T G \U \right).
\end{equation}
Averaged over $1024$ of such local variance maps, we find $\langle
\Delta \rangle \simeq 7.5 \, T G\U$ and blindly using the maximal
amplitude $\Delta = 117\,\muK$ gives $G\U \simeq 6 \times 10^{-6}$. As
mentioned above, this value of $\Delta$ is certainly underestimated
for the thawing strings. Indeed, instead of having one temperature
step of this amplitude over each $1.7^\circ$ circle, the local
variance maps essentially show that the delayed scaling network in the
thawing regime produces only one step of amplitude $8 T G\U$ over the
whole sky. Nevertheless, it is certainly fair to claim that values of
$G \U = \order{1} \times 10^{-6}$ may be considered as a minimum
threshold above which the string effects could be directly detectable
in the CMB maps.

\section{Induced power asymmetry}
\label{sec:mod}

Although thawing strings let relatively weak CMB imprints at small
angular scales, they genuinely generate a power asymmetry on the
largest angles. As can be seen in figure~\ref{fig:stgmaps2}, all maps
are associated with smooth gradients over the whole sky. As reported
in Refs.~\cite{Ade:2013nlj, Ade:2015sjc}, various large scale
anomalies measured in the Planck 2013 and 2015 data can be
phenomenologically described by a dipole modulation model. Such a
model has however been shown to have some significance on the largest
angular scales only~\cite{Quartin:2014yaa} and, in the following, we
quantify as much thawing strings could contribute to this effect.

\subsection{Dipole in local variance maps}

\begin{figure}
\begin{center}
\includegraphics[angle=90,width=\twofigw]{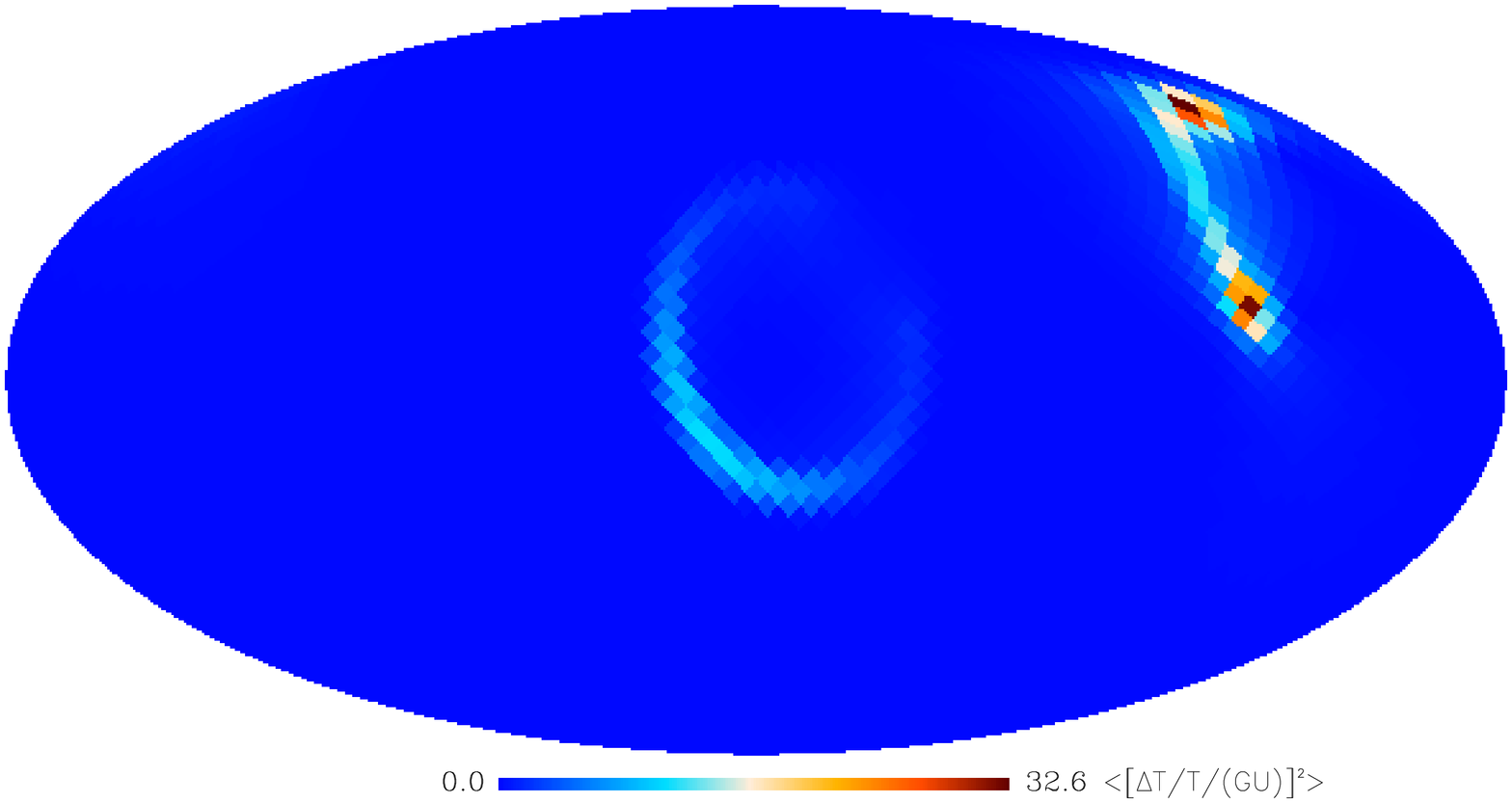}
\includegraphics[angle=90,width=\twofigw]{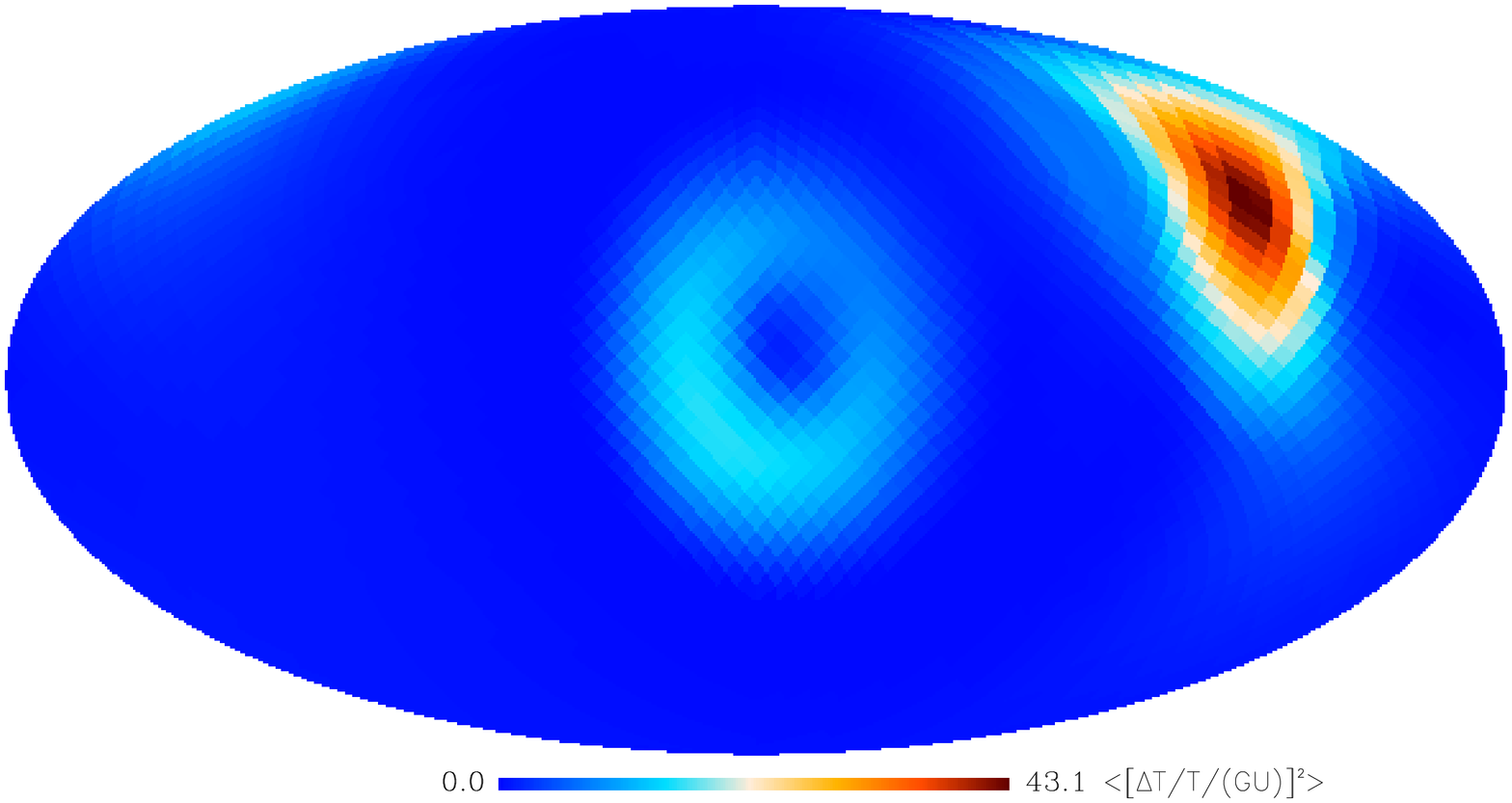}
\includegraphics[angle=90,width=\twofigw]{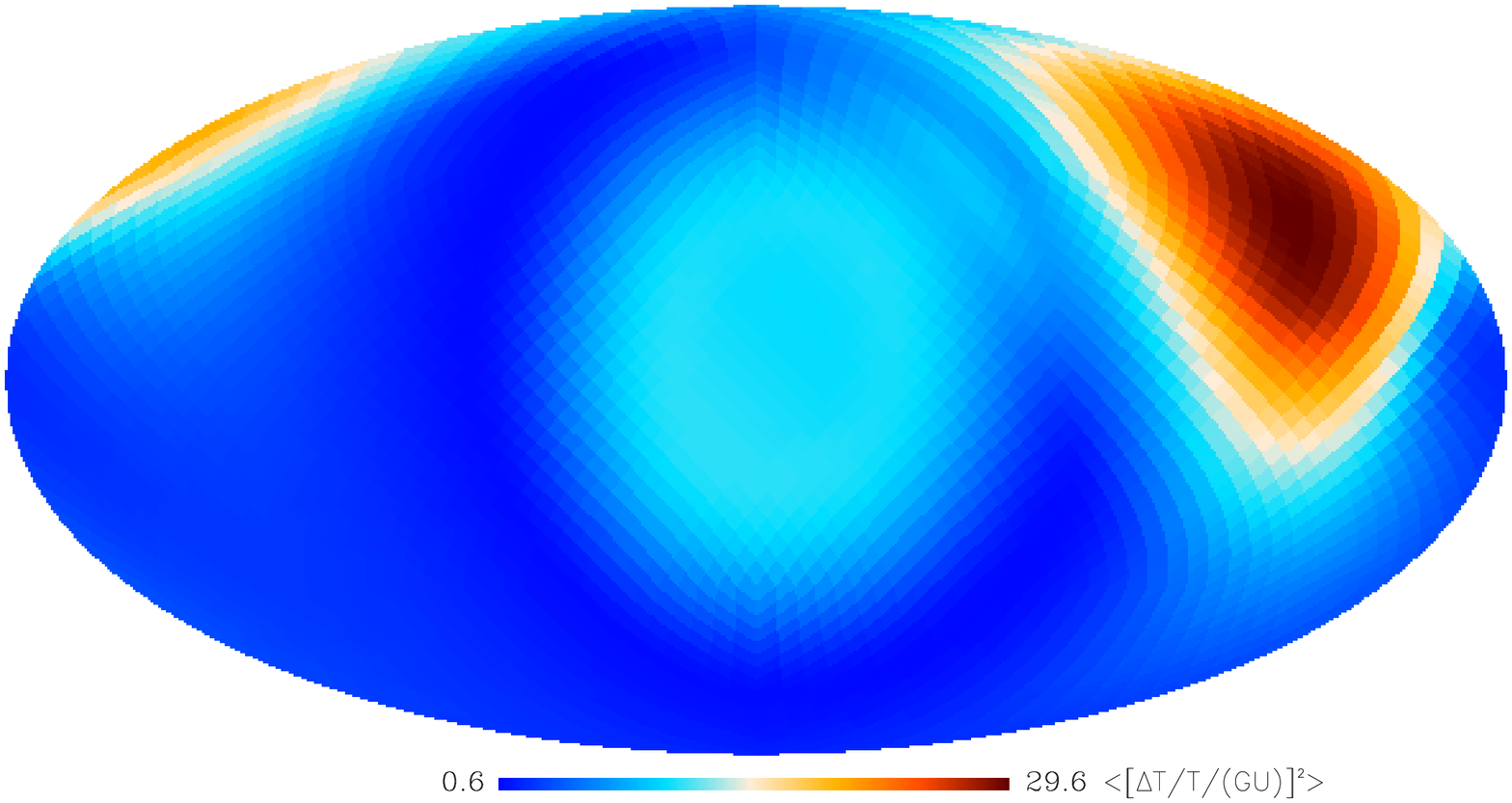}
\includegraphics[angle=90,width=\twofigw]{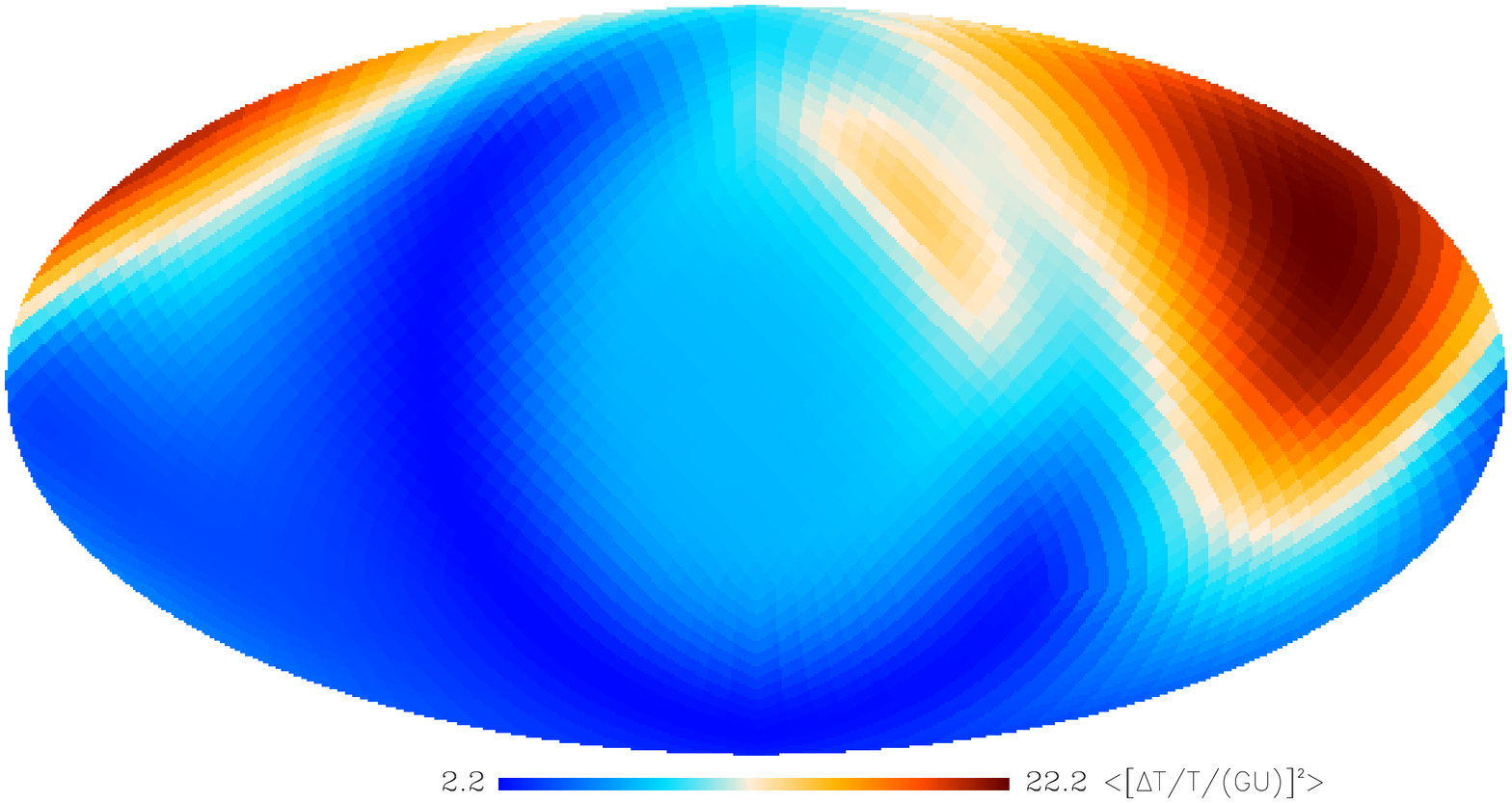}
\caption{Local variance maps obtained from the first CMB maps of
  figure~\ref{fig:stgmaps2} (top left). From top left to bottom right,
  the averaging angles shown are $6^\circ$, $24^\circ$, $48^\circ$ and
  $64^\circ$, respectively.}
\label{fig:varmaps2}
\end{center}
\end{figure}

As discussed in Ref.~\cite{Akrami:2014eta}, local variance maps of
temperature anisotropies over circles of given angular size can be
used to quantify the amplitude of a power asymmetry. In particular, a
pure dipole modulation along a given direction
$\unitw$~\cite{Akrami:2014eta, Gordon:2005ai}
\begin{equation}
\dfrac{\Delta T}{T}(\unitn) = (1+ A \unitn\cdot\unitw)
\left. \dfrac{\Delta T}{T} \right|_{\uLCDM}\,,
\label{eq:puredip}
\end{equation}
would also show up in the local variance map as a dipole of
amplitude $2A\sigma^2_{\uLCDM}$:
\begin{equation}
\sigma^2 = \sigma^2_{\uLCDM} + 2A  \sigma^2_{\uLCDM} \unitn\cdot
\unitw + \order{A^2}.
\end{equation}
As a result, one may test the existence of this anisotropy by fitting
both a monopole $\sigma^2_\umon$ and a dipole amplitude
$\sigma^2_\udip$ to the local variance maps. For a pure dipole
modulation as in Eq.~\eqref{eq:puredip}, the ratio $\sigma^2_\udip/(2
\sigma^2_\umon)\simeq A$ independently of the circle angular size.
For each of the $1024$ CMB maps generated in section~\ref{sec:simus},
we have constructed as many local variance maps by averaging over
circles of different angular size, namely $1.7^\circ$, $6^\circ$,
$12^\circ$, $24^\circ$, $36^\circ$, $48^\circ$ and $64^\circ$.

\begin{figure}
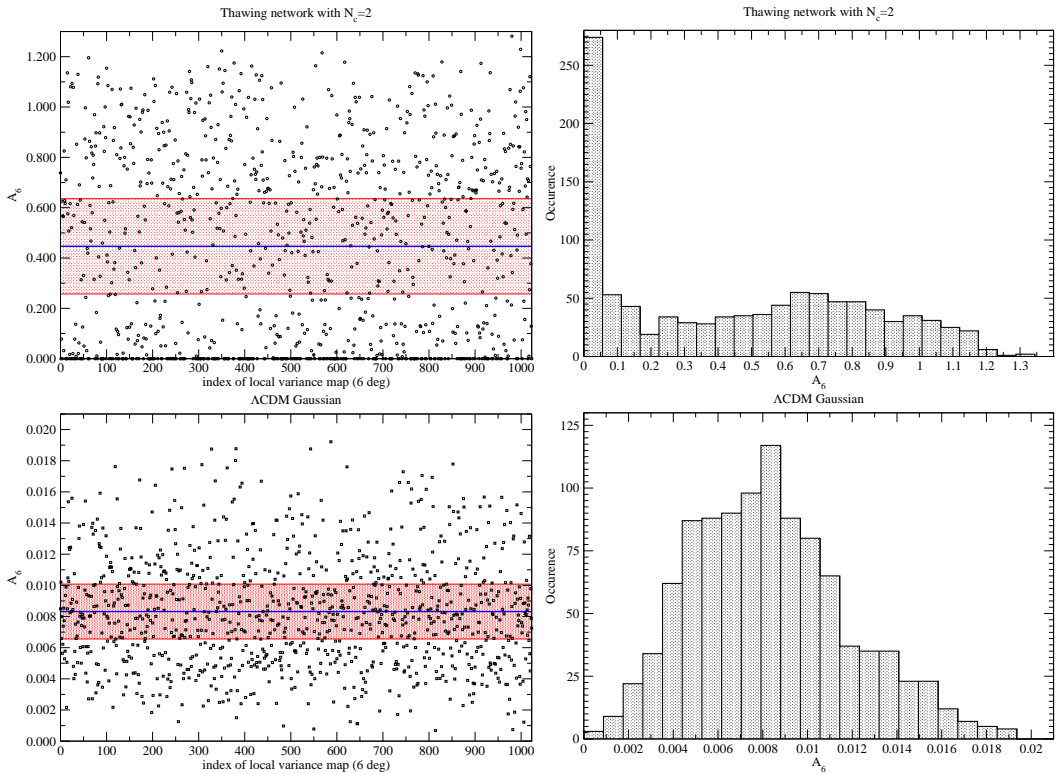

\begin{center}
\includegraphics[width=\twofigw]{stgmod_n2_ang6}
\includegraphics[width=0.965\twofigw]{stgdist_n2_ang6}
\includegraphics[width=\twofigw]{cmbmod_ang6}
\includegraphics[width=0.965\twofigw]{cmbdist_ang6}
\caption{The values of $A_6=\sigma^2_\udip/(2\sigma^2_\umon)$ obtained
  for $1024$ local variance maps of angular opening $6^\circ$ (left)
  and its distribution (right). The upper panels are for the thawing
  strings scenario with $\Ncorrini=2$ while the lower panels have been
  derived from pure $\LCDM$ CMB maps. The horizontal line corresponds
  to the mean value over $1024$ realizations while the shaded region
  represents the standard deviation. The dipole modulation for thawing
  strings can be two orders of magnitude larger than the one expected
  for Gaussian $\LCDM$.}
\label{fig:modulation}
\end{center}
\end{figure}

Figure~\ref{fig:varmaps2} illustrates the procedure by showing the
local variance maps associated with the top left realization of
figure~\ref{fig:stgmaps2} and for four angles. The local variance map
contains only a strong signal if a string is present within the
averaging angle. Therefore, for small angles, most of the sky
directions contain no strings and the local variance map exhibits a
strong signal only around the strings location.  On the contrary, for
larger angles, many more directions in the sky may contain a string
within the circle such that the string signal affects larger patches.

A monopole and a dipole are then fitted to each of these maps. The
direction of the dipole is random as it changes with each
realization. However, the ratio
$A_\theta=\sigma^2_\udip/(2\sigma^2_\umon)$ tells us how much the
strings may generate a signal looking like a dipole modulation at a
given angle $\theta$. In figure~\ref{fig:modulation}, we have plotted
the distribution of this ratio over $1024$ local variance maps
averaged over circles of $6^\circ$. Because a dipole modulation may
also appear within the ordinary temperature anisotropies of
inflationary origin, figure~\ref{fig:modulation} also shows the value
of $\sigma^2_\udip/(2\sigma^2_\umon)$ obtained from $1024$ random
realizations of a purely Gaussian $\LCDM$ anisotropies (also convolved
with a $10'$ Gaussian beam). These Gaussian CMB maps have been
generated using the {\healpix} library~\cite{Gorski:2004by} from an
angular power spectrum calculated by means of the {\camb}
code~\cite{Lewis:1999bs}. The cosmological parameter values have been
set according to the Planck 2015 favoured values~\cite{Adam:2015rua},
assuming a negligible contribution from tensor modes. For the
Gaussian $\LCDM$ realizations, one gets $\langle A_6 \rangle_{\uLCDM}
=8.3\times 10^{-3}$ whereas the strings with $\Ncorrini=2$ gives
$\langle A_6 \rangle_\udelay = 0.45$. As can be seen in
figure~\ref{fig:modulation}, the value of $\langle A_6
\rangle_\udelay$ is not very representative of the actual
distribution, this one being actually multivalued. There is a sharp
peak at vanishing value of $A_6$ while there is another local maximum
around $A_6 \simeq 0.7$. As mentioned before, this is because, for
$\Ncorrini=2$, around one quarter of the thawing string realizations
produces no string on our past light cone. Therefore, a more
representative value of the dipole modulation generically generated when
strings are actually present is $A_6 \simeq 0.7$; which is more than
two orders of magnitude greater than the Gaussian expected value.

\subsection{Averaging angle dependency}

\begin{figure}
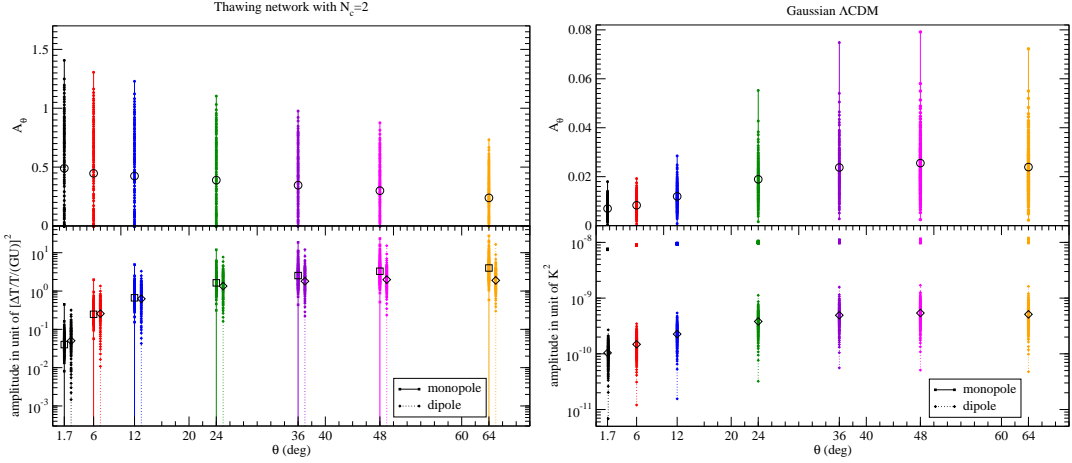

\begin{center}
\includegraphics[width=\twofigw]{stgmod_n2_all}
\includegraphics[width=\twofigw]{cmbmod_all}
\caption{Dependency of the dipole modulation amplitude
  $A_\theta=\sigma^2_\udip/(2 \sigma^2_\umon)$ with respect to the
  averaging angle $\theta$ used to construct the local variance
  maps. The left panels are for thawing strings and can be compared to
  the expected pure Gaussian $\LCDM$ realizations plotted on the right
  panels. Each point at a given $\theta$ represents one of the $1024$
  realizations while the largest black circle is the mean value. The
  lower panels show the absolute value of the monopole
  ($\sigma^2_\umon$) and dipole ($\sigma^2_\udip$) values (the string
  dipole values have been slightly shifted to the right for clarity).}
\label{fig:avgangle2}
\end{center}
\end{figure}

\begin{table}
\begin{center}
\begin{tabular}{|c|c|c|c|c|c|c|c|}
\hline
$\theta$ & $1.7^\circ$ & $6^\circ$ & $12^\circ$ & $24^\circ$ &
  $36^\circ$ & $48^\circ$ & $64^\circ$ \\ \hline
$\langle A_\theta \rangle_\udelay$ & $0.49$ & $0.45$ & $0.42$ & $0.39$ & $0.35$
& $0.30$ & $0.24$  \\ \hline
$\langle \sigma^2_\udip \rangle_\udelay/(T G\U)^2$ & $0.05$ & $0.26$ & $0.63$ &
$1.34$ & $1.80$ & $1.98$ & $ 1.90$ \\ \hline
$\langle \sigma^2_\umon \rangle_\udelay/(TG\U)^2$ & 0.04 & 0.25 & 0.67
& 1.63 & 2.53& 3.28 & 3.99 \\ \hline
\end{tabular}
\caption{Mean values of $A_\theta$, $\sigma^2_\umon$ and
  $\sigma^2_\udip$ obtained over $1024$ realizations of the thawing
  string networks with $\Ncorrini=2$ (see also
  figure~\ref{fig:avgangle2}).}
\label{tab:mean2}
\end{center}
\end{table}

\begin{table}
\begin{center}
\begin{tabular}{|c|c|c|c|c|c|c|c|}
\hline
$\theta$ & $1.7^\circ$ & $6^\circ$ & $12^\circ$ & $24^\circ$ &
  $36^\circ$ & $48^\circ$ & $64^\circ$ \\ \hline
$\langle A_\theta \rangle_\uLCDM$ & 0.007& 0.008 & 0.012 & 0.019 &
0.024 & 0.025 & 0.024  \\ \hline
$10^{-2} \langle \sigma^2_\udip \rangle_\uLCDM$ ($\muK^2$) & $1.04$ &
$1.48$ & $2.26$ & $3.80$ & $4.93$ & $5.39$ & $5.11$ \\ \hline
$10^{-4} \langle \sigma^2_\umon \rangle_\uLCDM$ ($\muK^2$) & 0.75 & 0.89
& 0.95 & 1.00 & 1.03 & 1.05 & 1.06 \\ \hline
\end{tabular}
\caption{Mean values of $A_\theta$, $\sigma^2_\umon$ and
  $\sigma^2_\udip$ obtained over $1024$ realizations of Gaussian
  $\LCDM$ maps (see also figure~\ref{fig:avgangle2}).}
\label{tab:meancdm}
\end{center}
\end{table}

Figure~\ref{fig:avgangle2} shows the dependency of $A_\theta$ with
respect to the local variance map averaging angle $\theta$, for both
the thawing strings scenario with $\Ncorrini=2$ (left panel) and the
pure $\LCDM$ model (right panel). There is a weak dependence of
$A_\theta$ with respect to $\theta$ for both the strings and $\LCDM$,
with the notable difference that the distribution of $A_\theta$
slightly goes down for strings whereas it increases for $\LCDM$ for
larger values of $\theta$. For essentially all angles, $A_\theta$
generated by strings is two orders of magnitude greater than the
Gaussian one. Moreover, the distribution of $A_\theta$ has a large
variance, over $1024$ realizations the maximum value may be larger
than unity while the non-vanishing value for the minimum may be as low
as $\order{1} \times 10^{-3}$. For instance, for $\theta=6^\circ$, one
gets $\min\{A_6 | A_6 \ne 0\} = 4 \times 10^{-3}$. Let us notice that
the values obtained here for $\LCDM$ (right panels) are close to the
ones derived from the Planck Full Focal Plane isotropic simulations
(FFP6) of Ref.~\cite{Akrami:2014eta}. This agreement might be
surprising at first, as our simulations are pure Gaussian $\LCDM$
without any instrumental noise, but, at these scales of interest, most
of the variance is indeed generated by the CMB.

The lower panels of figure~\ref{fig:avgangle2} display the absolute
value of the fitted monopole $\sigma^2_\umon$ and dipole
$\sigma^2_\udip$ with respect to $\theta$. As opposed to the $\LCDM$
case, in which both of these values increase only by a small factor
with $\theta$, the delayed scaling string variance maps exhibit an
increase by almost two orders of magnitude from $\theta=1.7^\circ$ to
$64^\circ$. Such an effect results from both the actual string induced
temperature patterns together with the construction of the local
variance map. As can be seen in figure~\ref{fig:varmaps2}, the number
of sky directions which are sensitive to the presence of a localized
temperature pattern obviously increases with the opening angle
$\theta$. This effect does not change very much the amplitude (and
direction) of the modulation but boosts the averaged value of both the
monopole and dipole simultaneously. On the contrary, because Gaussian
$\LCDM$ fluctuations generate the same variance over all directions,
changing the opening angle $\theta$ does not fundamentally modify the
mean value of the monopole and dipole.

Qualitatively, the previous results suggest that if a few thawing
strings actually contribute to the CMB sky, in addition to the
Gaussian $\LCDM$ anisotropies, they can produce a dipole modulation in
the local variance map. The lower panels of figure~\ref{fig:avgangle2}
show that a typical signature is that such a modulation should be
detectable only on the large angular scales.

\subsection{Mixture of string-induced and Gaussian anisotropies}
\label{sec:mixture}

\begin{figure}
\begin{center}
\includegraphics[width=\twofigw]{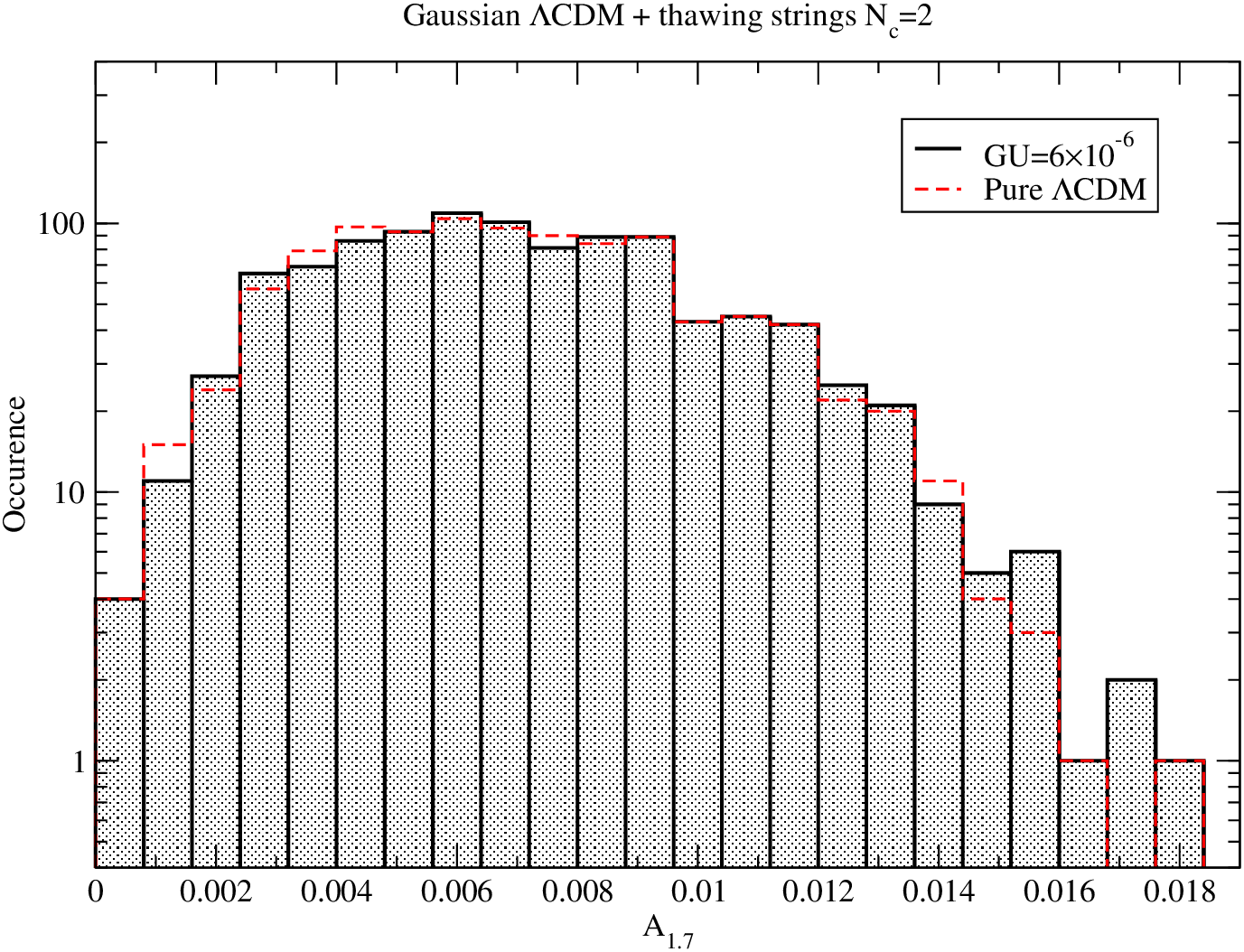}
\includegraphics[width=\twofigw]{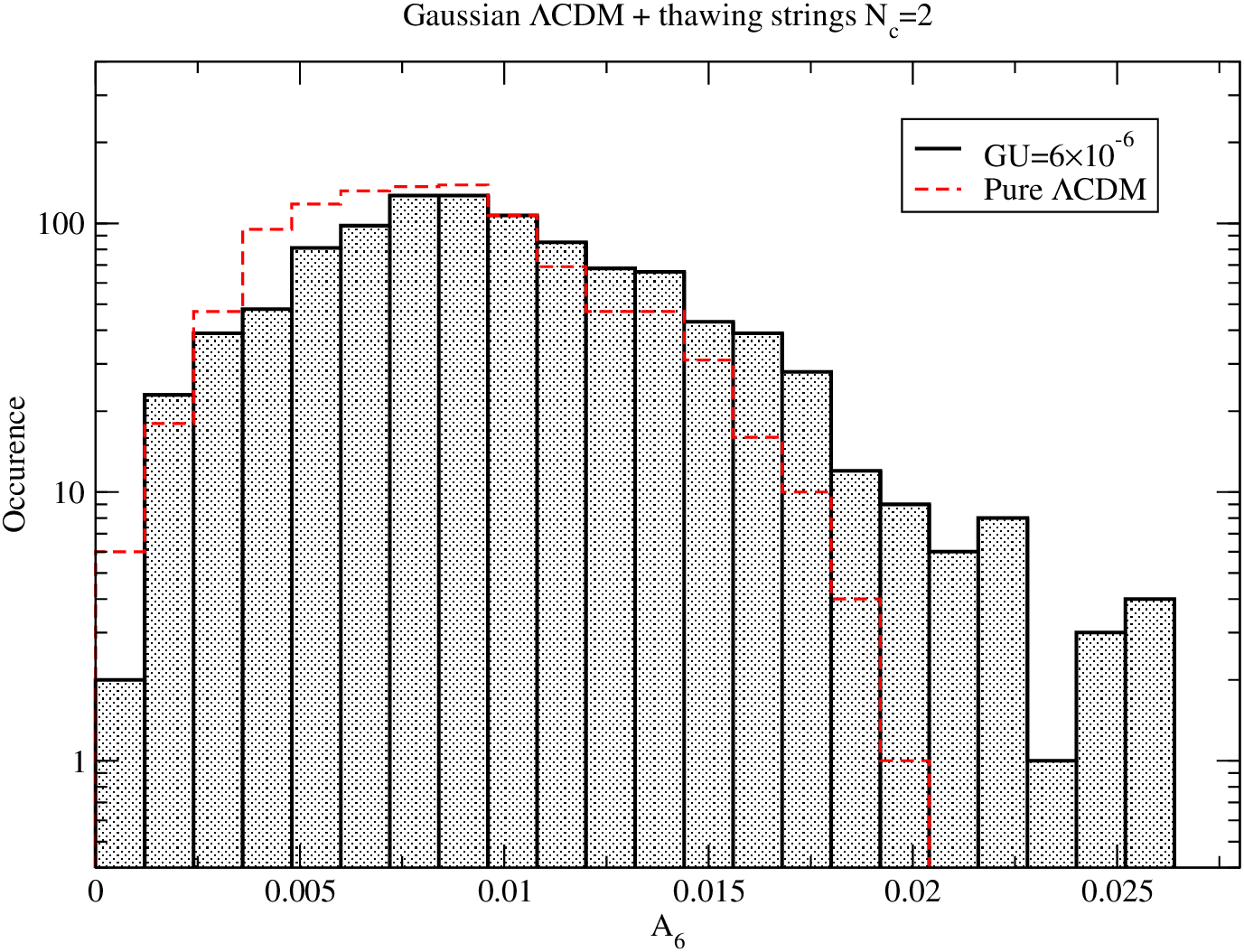}
\includegraphics[width=\twofigw]{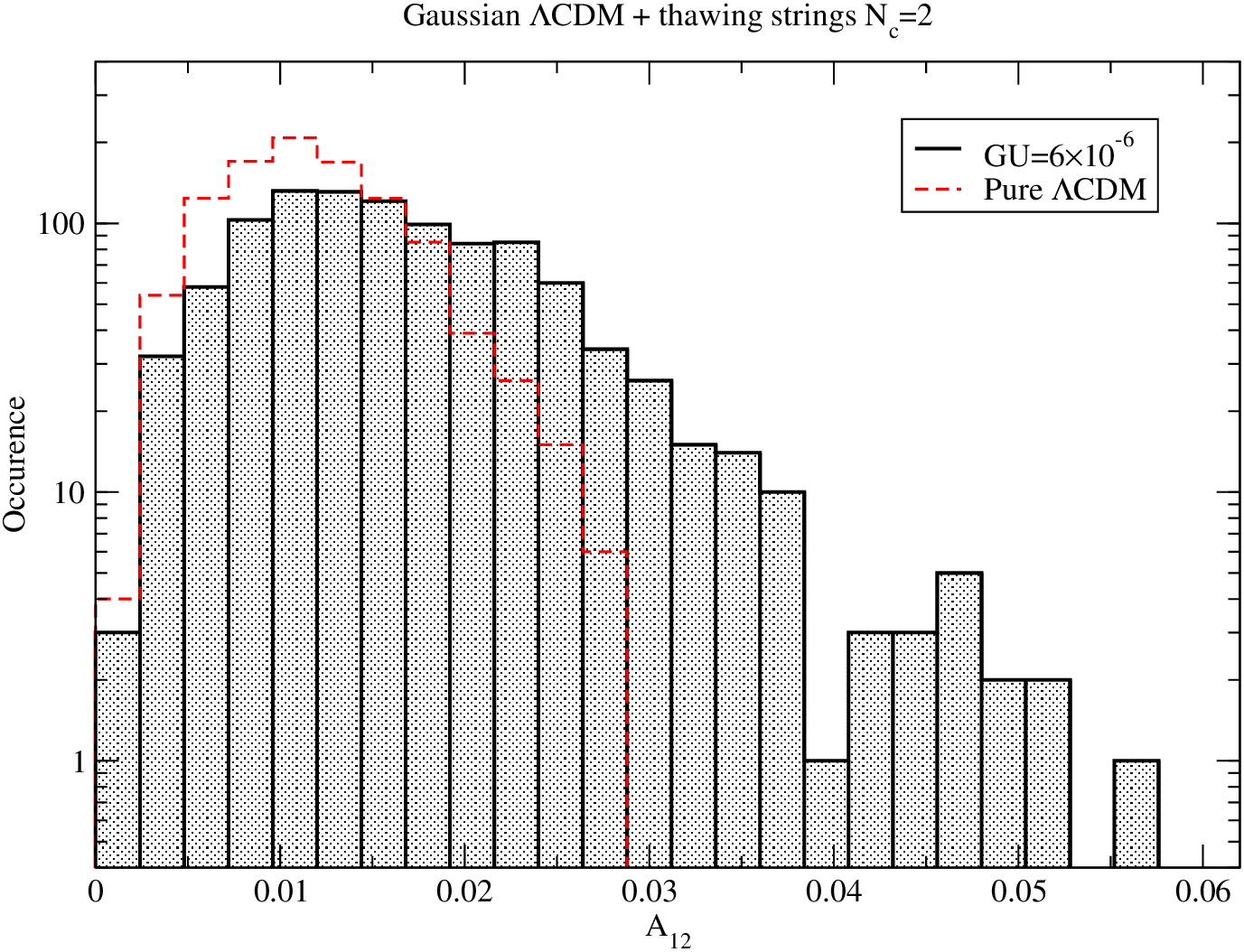}
\includegraphics[width=\twofigw]{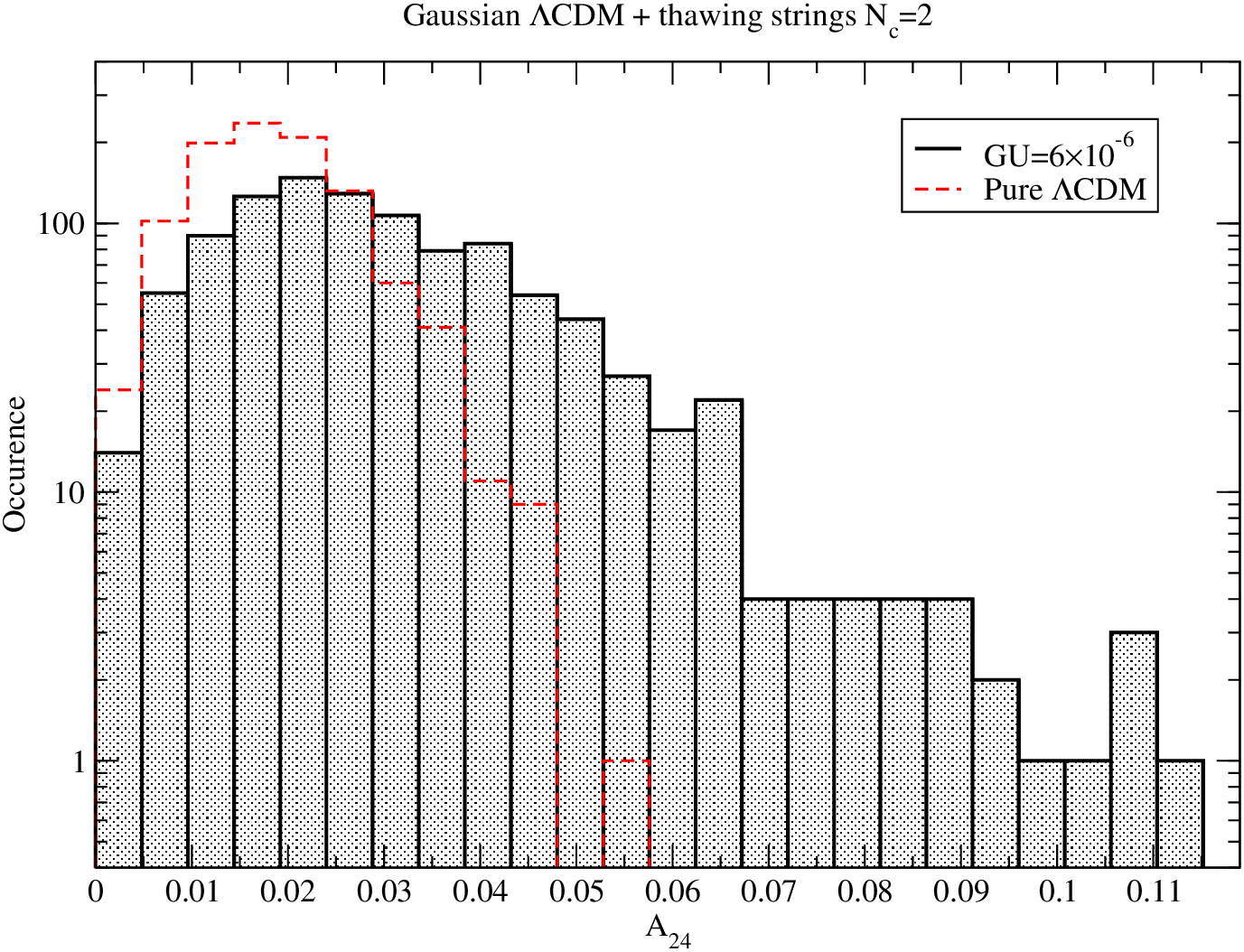}
\includegraphics[width=\twofigw]{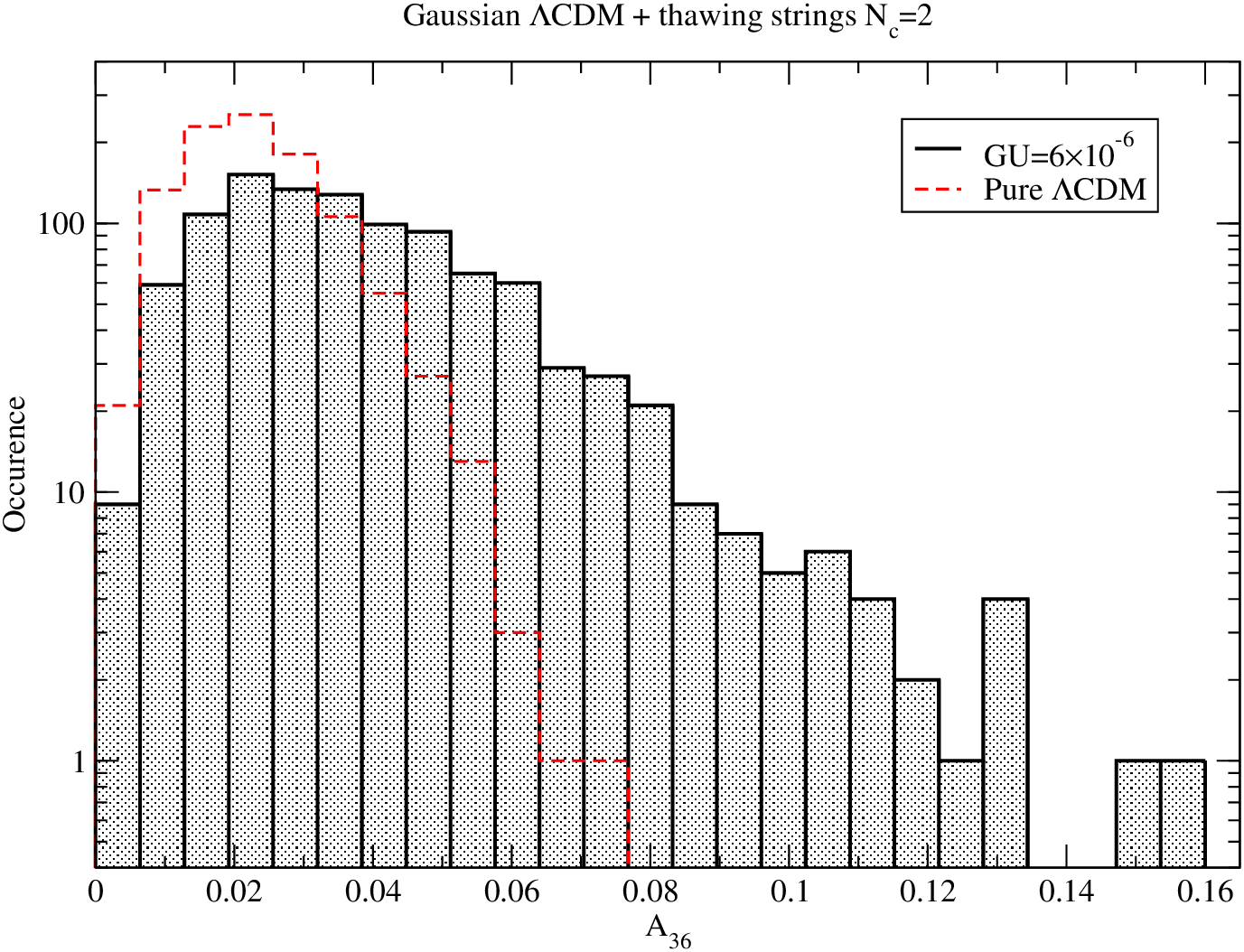}
\includegraphics[width=\twofigw]{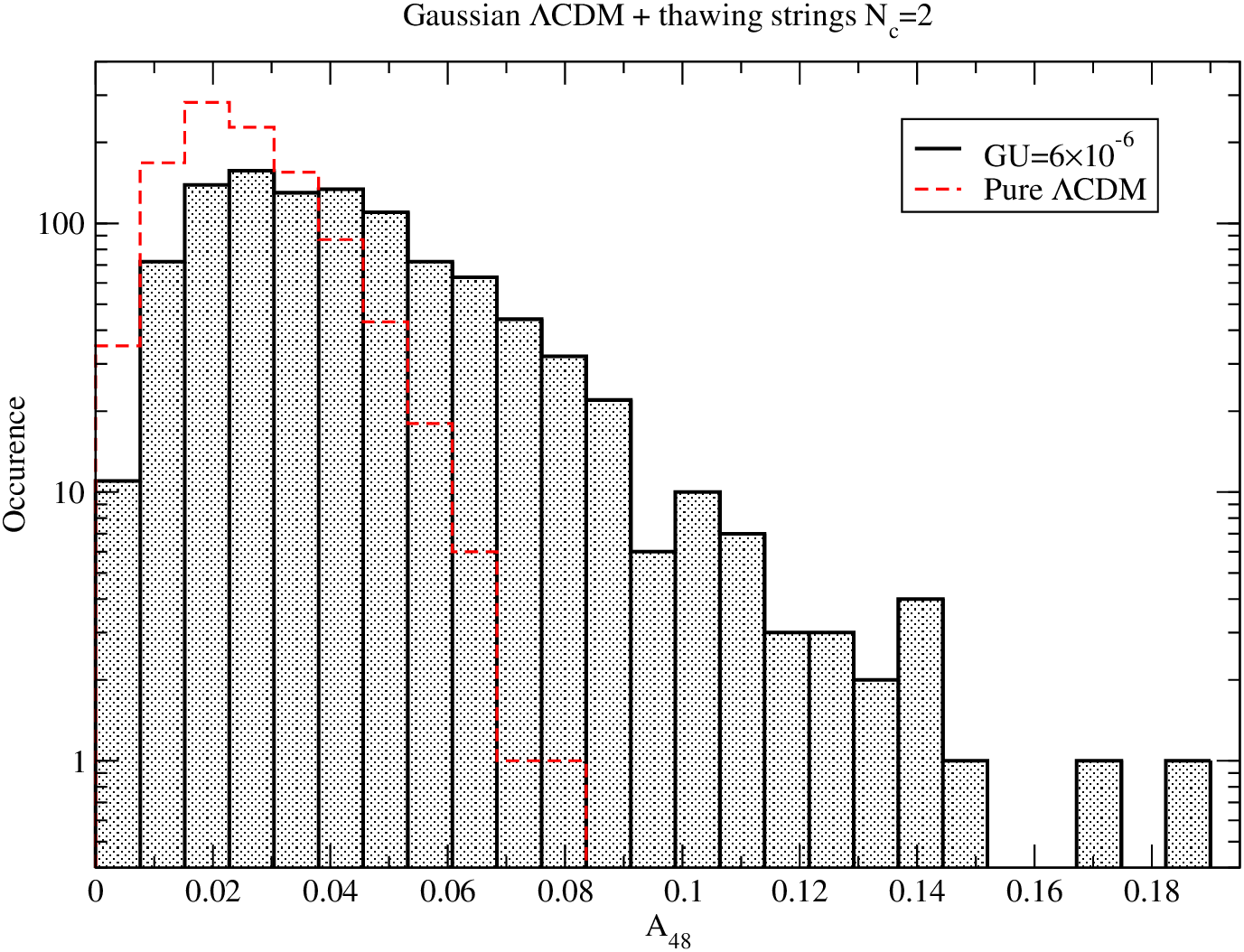}
\caption{Distribution of the dipole modulation amplitude $A_\theta$
  obtained for the opening angles $\theta=1.7^\circ$,
  $\theta=6^\circ$, $\theta=12^\circ$, $\theta=24^\circ$,
  $\theta=36^\circ$ and $\theta=48^\circ$ from a mixture of Gaussian
  $\LCDM$ fluctuations and thawing strings with $\Ncorrini=2$ and
  $G\U=6\times 10^{-6}$. The pure Gaussian $\LCDM$ distributions have
  been reported for comparison. For angles $\theta \le 2^\circ$, there
  are essentially no differences with respect to pure Gaussian
  anisotropies whereas significantly larger dipole amplitudes appear
  for $\theta \ge 6^\circ$.}
\label{fig:n2_GU6EM6}
\end{center}
\end{figure}

\subsubsection{Mean values}

We now consider that a few strings actually contribute to the overall
CMB anisotropies, i.e., $\delta T = \delta T_\uLCDM + \delta T_\udelay$
in which both contributions are assumed to be
uncorrelated\footnote{Such an assumption may no longer be true in some
  particular scenarios~\cite{Tseng:2009xw}.}. Denoting by $\mon \equiv
\sigma^2_\umon$ and $\dip \equiv \sigma^2_\udip$ one gets
\begin{equation}
\begin{aligned}
A & = \dfrac{\dip_\uLCDM + \dip_\udelay}{2(\mon_\uLCDM +
  \mon_\udelay)} \simeq \left(1-\dfrac{\mon_\udelay}{\mon_\uLCDM} \right)
A_\uLCDM + \dfrac{\mon_\udelay}{\mon_\uLCDM} A_\udelay, \\
& \simeq \left[1-\dfrac{\bar{\mon}_\udelay}{\bar{\mon}_\uLCDM} (G\U)^2
  \right] A_\uLCDM + \dfrac{\bar{\mon}_\udelay}{\bar{\mon}_\uLCDM}
(G\U)^2 A_\udelay,
\end{aligned}
\label{eq:modmix}
\end{equation}
where we have kept only the leading order terms in
$\mon_\udelay/\mon_\uLCDM$ in the first line and
introduced the dimensionless quantities
\begin{equation}
\bar{\mon}_\udelay \equiv \dfrac{\mon_\udelay}{(TG\U)^2}\,, \qquad
\bar{\mon}_\uLCDM \equiv \dfrac{\mon_\uLCDM}{T^2}\,.
\end{equation}
The second line of Eq.~\eqref{eq:modmix} renders explicit the
dependency in $G\U$ and shows that the string contribution may boost
the dipole modulation amplitude. For instance, matching a given dipole
modulation amplitude $A_\uobs$, one obtains
\begin{equation}
G\U \simeq \sqrt{\dfrac{\bar{\mon}_\uLCDM}{\bar{\mon}_\udelay}}
\sqrt{\dfrac{A_\uobs - A_\uLCDM}{A_\udelay - A_\uLCDM}}\,,
\label{eq:GUavg}
\end{equation}
an expression valid only for $A_\uobs>A_\uLCDM$ and $A_\udelay >
A_\uLCDM$, which is the situation we are interested in. Taking a
fiducial value $A_\uobs \simeq 0.06$ compatible with the one reported
by the Planck collaboration~\cite{Ade:2013nlj, Ade:2015sjc}, and
blindly using mean values from tables~\ref{tab:mean2} and
\ref{tab:meancdm}, one gets $G\U \simeq 7.7 \times 10^{-6}$ for
$\theta=64^\circ$, $G\U \simeq 7.7 \times 10^{-6}$ for
$\theta=48^\circ$, $G\U \simeq 6.6 \times 10^{-6}$ for $\theta =
36^\circ$, $G\U \simeq 9.5 \times 10^{-6}$ for $\theta = 24^\circ$,
$G\U \simeq 1.5 \times 10^{-5}$ for $\theta = 12^\circ$, $G\U \simeq
2.4 \times 10^{-5}$ for $\theta=6^\circ$ and $G\U \simeq 5.2 \times
10^{-5}$ for $\theta=1.7^\circ$. These figures confirm that, for a
given value of $G\U=\order{1}\times 10^{-6}$, the strings may generate
a dipole modulation compatible with the one measured, but solely on
the large angular scales. Interestingly, these values are quite
natural for GUT scale cosmic strings and are of the same order of
magnitude as the ones quoted in
section~\ref{sec:directobs}. Therefore, if delayed scaling strings in
the thawing regime are the underlying cause of the dipole modulation,
the (very few) associated temperature discontinuities should be
detectable in the CMB sky, provided they are not masked by
foregrounds.

\subsubsection{Local variance maps}

The values of $G\U$ quoted before have been estimated by using the
mean values of table~\ref{tab:mean2}, which may not be representative
of the actual distribution. As can be checked in
figure~\ref{fig:avgangle2}, the mean value of $\langle A_\theta
\rangle_\udelay$ underestimates the actual distribution of
non-vanishing realizations. This effect can be taken into account by
directly mixing the string and $\LCDM$ simulated CMB maps. In
figure~\ref{fig:n2_GU6EM6}, we have represented the actual
distribution of $A_\theta$ obtained over $1024$ local variance maps
created from a mixture of random Gaussian $\LCDM$ anisotropies with
random thawing string configurations having $\Ncorrini=2$ and
$G\U=6\times 10^{-6}$. These distributions show that, for such a value
of $G\U$, there is no significant deviation with respect to the
$\LCDM$ Gaussian distribution at small angles, $\theta \le
2^\circ$. On the contrary, already for $\theta \ge 6^\circ$, the
distribution of possible dipole modulation amplitudes in the mixed
scenario exhibits a long tail towards large values. For
$\theta=12^\circ$, amplitude larger than $0.06$ remains rare but far
more probable than in the pure Gaussian case whereas for $\theta \ge
24^\circ$ they appear in $5\%$ of the $1024$ realizations. These
numbers accordingly increases with the value of $G\U$.

\subsection{Discussion}

In this section, we have shown that the thawing string network with
$\Ncorrini=2$ genuinely generates a large scale dipole modulation in
the local variance maps. The string tension required for matching the
currently observed value is of the order $G\U = \order{1} \times
10^{-6}$, which suggests that the temperature patterns generated by
these very few strings may be detectable with dedicated
searches. Another parameter which affects the signature of a delayed
string network in the CMB is the correlation length
$\xi/\eta|_\ulss$. As discussed before, for the value
$\xi/\eta_\ulss\simeq 32$ used here ($\Ncorrini=2$), around a quarter
of the realizations do not exhibit any strings on our past light cone
and one may wonder what would be the situation with smaller initial
correlation lengths. In the context of inflation, the strings would be
formed later on.

\section{Cold spots from thawing loops}
\label{sec:thawing}

Decreasing the correlation length of the string network at last
scattering increases the number of strings on our past light cone and
render the network properties closer to the delayed scaling scenario
studied in Ref.~\cite{Kamada:2014qta}. As illustrated in
figure~\ref{fig:stgcls24}, for $\Ncorrini=4$, the angular power
spectrum is larger than for $\Ncorrini=2$. At $\ell=10$, one has
$C_{10}^{\Ncorrini=4}/ C_{10}^{\Ncorrini=2} \simeq 4.6 $ and at
$\ell=400$, $C_{400}^{\Ncorrini=4}/ C_{400}^{\Ncorrini=2} \simeq
6.0$. For values of $\xi$ smaller than the Hubble radius at last
scattering, one would recover the configuration of a scaling network,
having a large power spectrum and strongly constrained values of
$G\U$. Moreover, by the central limit theorem, one expects the
properties of the dipole modulation to follow more closely a Gaussian
statistics.

By decreasing $\xi$ from large super-Hubble values to Hubble-like
values, one expects more thawing loops to be visible. Their typical
behaviour is to shrink under their tension and their dynamics becomes
close to the Nambu-Goto evolution in Minkowski space. The GKS effect
on the CMB necessarily yields a cooler region inside the loop. Such a
situation already occurs for $\Ncorrini=2$ as can be seen in the first
panel of figure~\ref{fig:stgmaps2}. If one increases $\Ncorrini$, the
angular size of the shrinking loops becomes smaller, their velocity
higher and these events could provide a natural explanation of the
so-called ``cold spot''~\cite{Cruz:2006sv, Das:2008es, Cruz:2008sb,
  Finelli:2014yha}.

\begin{figure}
\begin{center}
\includegraphics[angle=90,width=\twofigw]{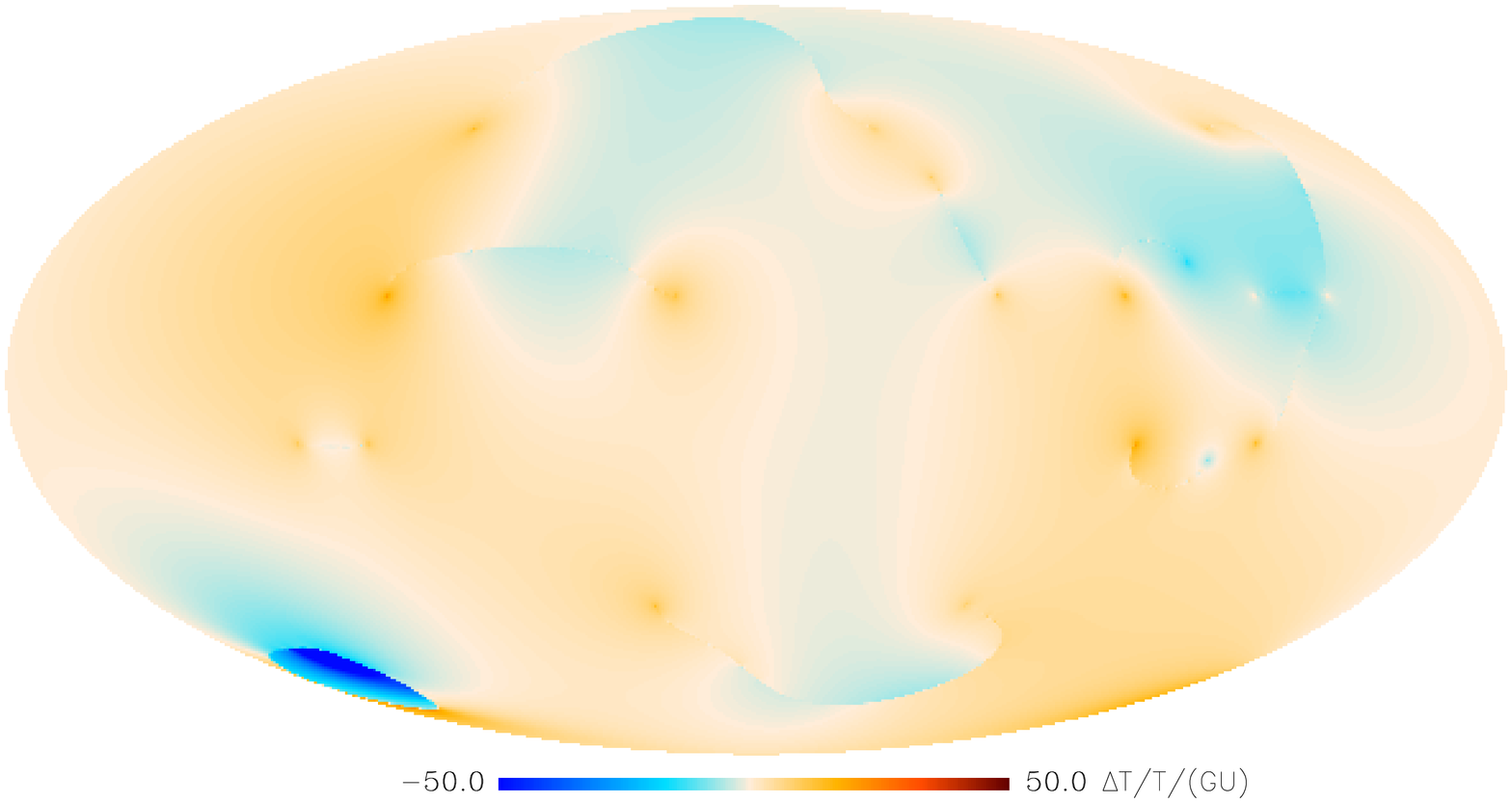}
\includegraphics[angle=90,width=\twofigw]{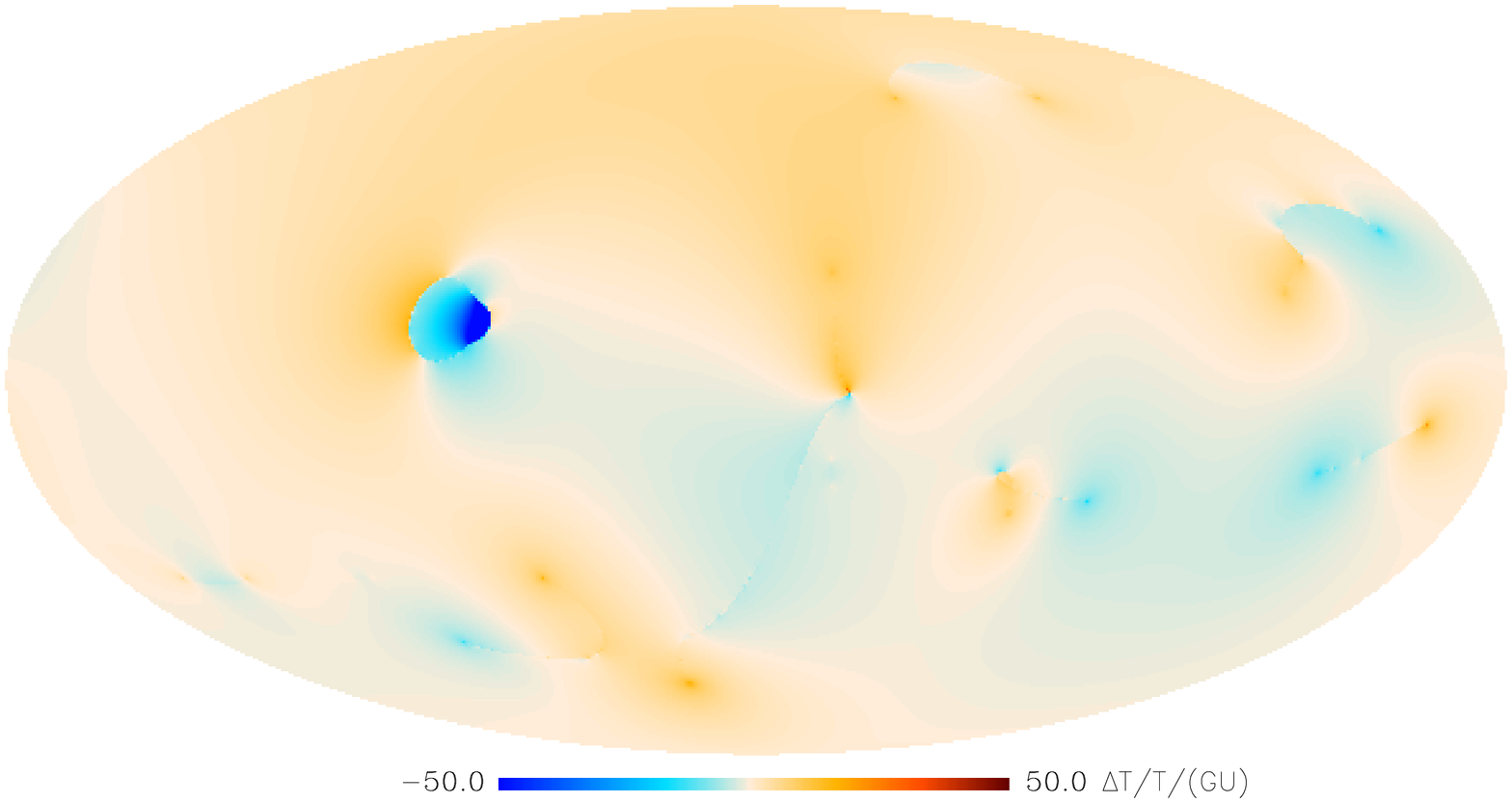}
\includegraphics[angle=90,width=\twofigw]{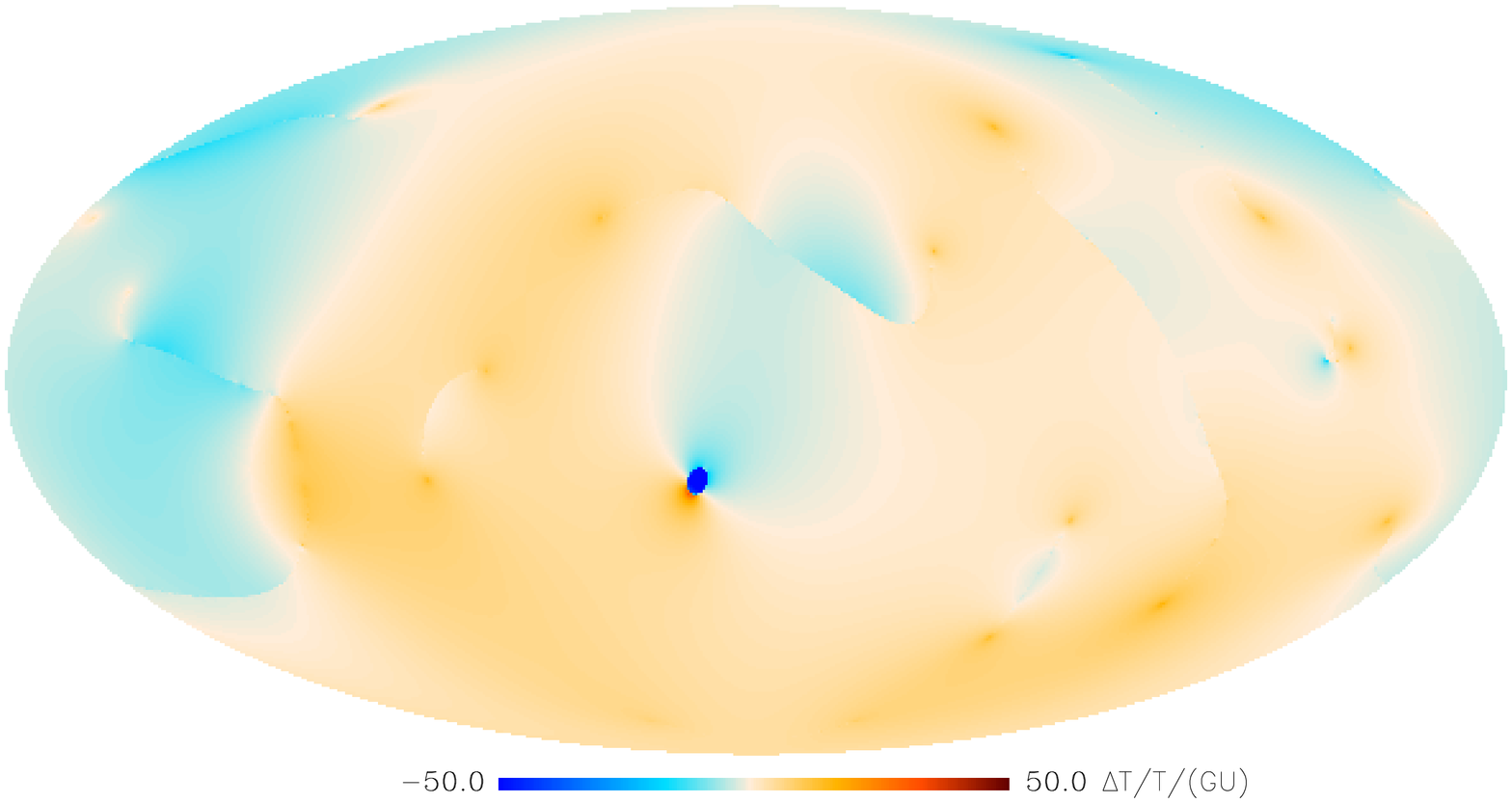}
\includegraphics[angle=90,width=\twofigw]{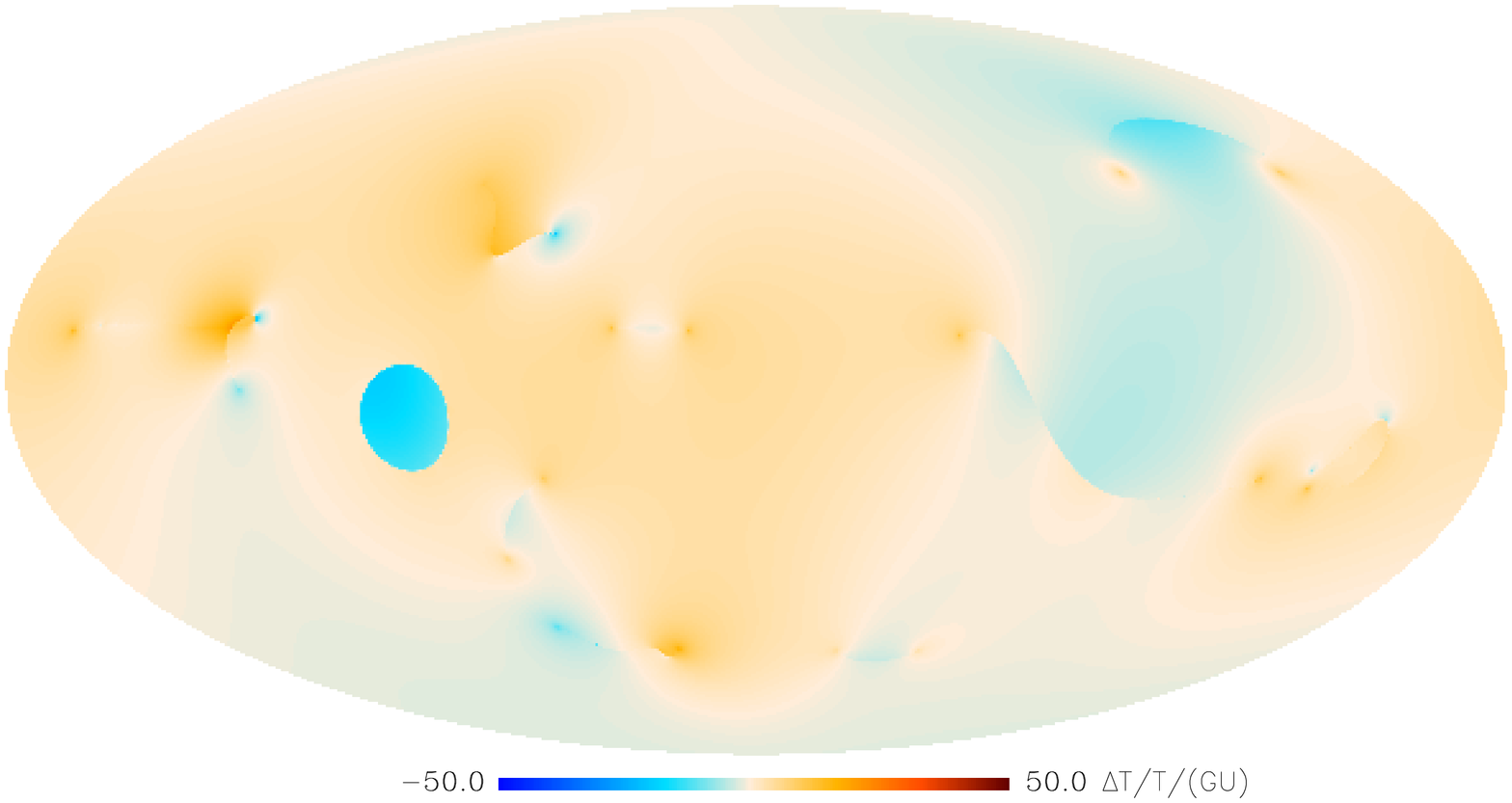}
\caption{Four realizations of the CMB sky temperature generated by
  delayed scaling strings in the thawing regime and having an initial
  correlation length $\xi = 16 \eta_\ulss$ ($\Ncorrini=4$). They
  exhibit one thawing loop generating a cold spot.}
\label{fig:stgmaps4}
\end{center}
\end{figure}

\begin{figure}
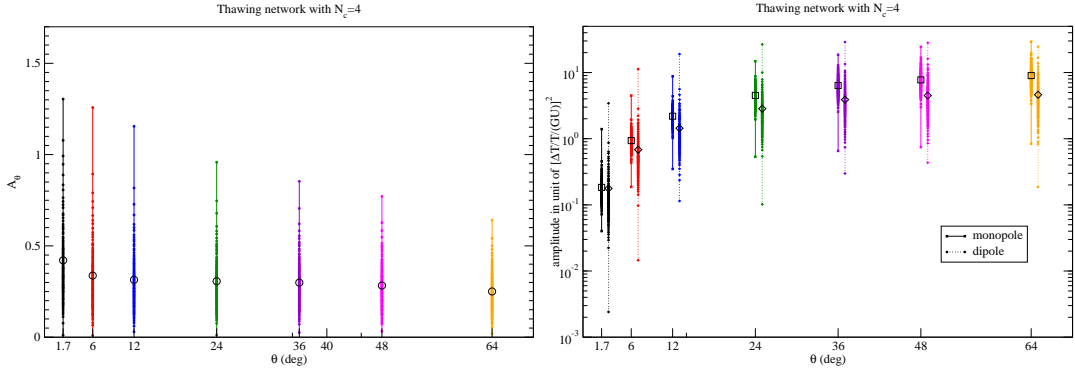

\begin{center}
\includegraphics[width=\twofigw]{stgmod_n4_all_A}
\includegraphics[width=\twofigw]{stgmod_n4_all_mondip}
\caption{Modulation amplitude $A_\theta$, monopole $\sigma^2_\umon$
  and dipole $\sigma^2_\udip$ in the local variance maps of thawing
  strings with $\Ncorrini=4$ (to be compared with
  figure~\ref{fig:avgangle2}). The mean values are represented with a
  bigger symbol and have been reported in table~\ref{tab:mean4}.}
\label{fig:avgangle4}
\end{center}
\end{figure}

\begin{table}
\begin{center}
\begin{tabular}{|c|c|c|c|c|c|c|c|}
\hline
$\theta$ & $1.7^\circ$ & $6^\circ$ & $12^\circ$ & $24^\circ$ &
  $36^\circ$ & $48^\circ$ & $64^\circ$ \\ \hline
$\langle A_\theta \rangle_\udelay$ & $0.42$ & $0.34$ & $0.31$ & $0.31$ & $0.30$
& $0.28$ & $0.25$  \\ \hline
$\langle \sigma^2_\udip \rangle_\udelay/(T G\U)^2$ & $0.18$ & $0.68$ & $1.44$ &
$2.84$ & $3.89$ & $4.49$ & $ 4.62$ \\ \hline
$\langle \sigma^2_\umon \rangle_\udelay/(TG\U)^2$ & 0.18 & 0.93 & 2.19
& 4.51 & 6.37 & 7.75 & 8.99 \\ \hline
\end{tabular}
\caption{Mean values of $A_\theta$, $\sigma^2_\umon$ and
  $\sigma^2_\udip$ obtained over $1024$ realizations of the thawing
  string networks with $\Ncorrini=4$.}
\label{tab:mean4}
\end{center}
\end{table}

We have generated $1024$ CMB maps associated with a thawing string
network with $\Ncorrini=4$ and thus having $\xi/\eta|_\ulss \simeq
16$. A few maps have been represented in figure~\ref{fig:stgmaps4} and
show the presence of one thawing loop, at most. From these maps, we
have derived the local variance maps averaged over various angles as
in the previous section. The amplitude of the dipole modulation is
still very large compared to the Gaussian $\LCDM$ value. The
distribution and mean values of $A_\theta$, $\sigma^2_\umon$ and
$\sigma^2_\udip$ for $\Ncorrini=4$ have been plotted in
figure~\ref{fig:avgangle4} and reported in table~\ref{tab:mean4}. From
Eq.~\eqref{eq:GUavg}, the typical value of $G\U$ required to get
$A_\uobs\simeq 0.06$ is $G\U \simeq 5\times 10^{-6}$ for
$\theta=64^\circ$, $G\U \simeq 5 \times 10^{-6}$ for
$\theta=48^\circ$, $G\U \simeq 5.3 \times 10^{-6}$ for
$\theta=36^\circ$, $G\U \simeq 6.5 \times 10^{-6}$ for
$\theta=24^\circ$, $G\U \simeq 9.7 \times 10^{-7}$ for
$\theta=12^\circ$, $G\U \simeq 1.4 \times 10^{-5}$ for
$\theta=6^\circ$ and $G\U \simeq 2.7 \times 10^{-5}$ for
$\theta=1.7^\circ$. These numbers are slightly less than the ones
obtained with $\Ncorrini=2$ which may be attributed to the higher
number of strings visible and the corresponding larger value of the
mean variance (see table~\ref{tab:mean4}).

\begin{figure}
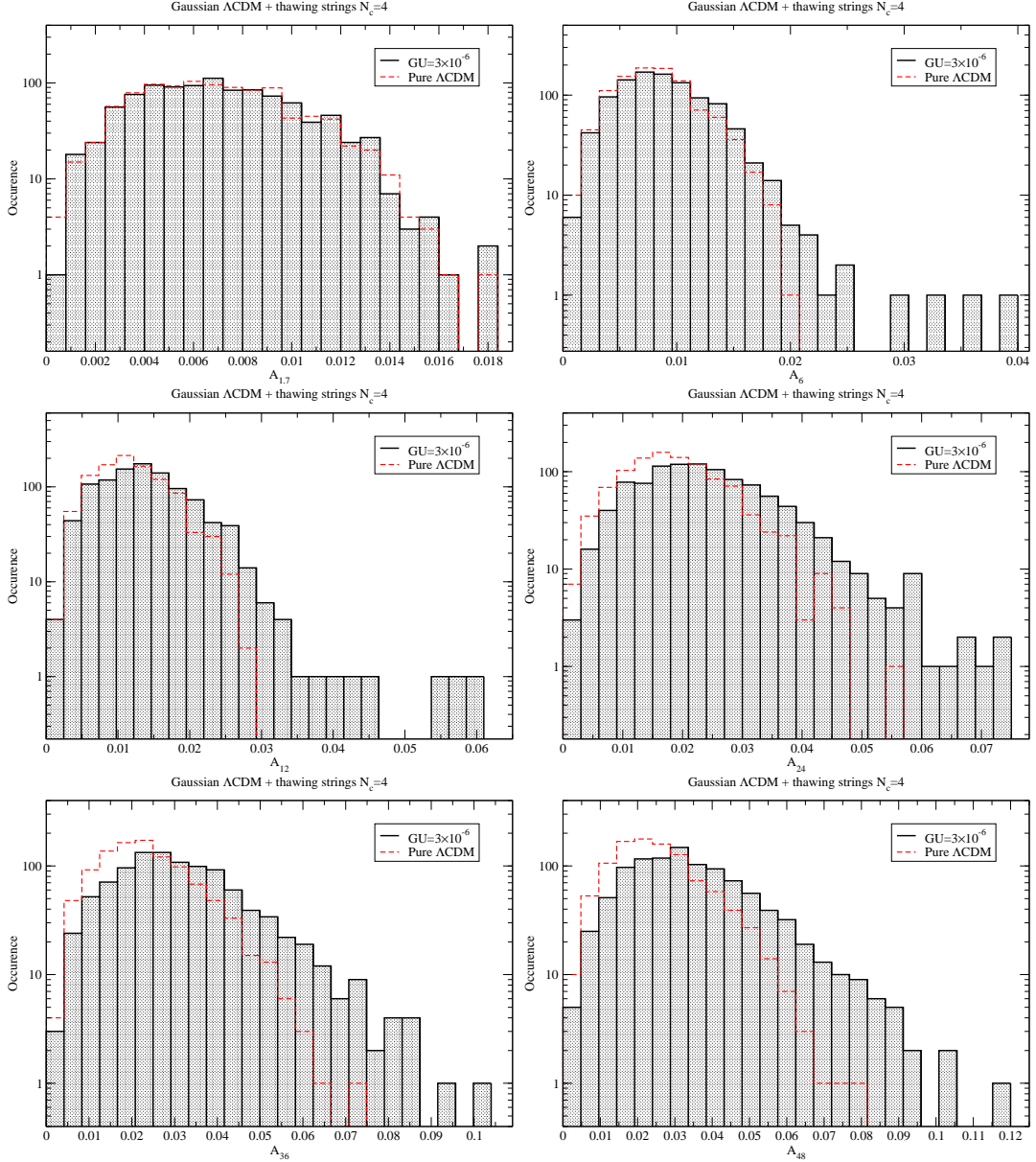

\begin{center}
\includegraphics[width=\twofigw]{mixdist_n4_gu3em6_ang2}
\includegraphics[width=\twofigw]{mixdist_n4_gu3em6_ang6}
\includegraphics[width=\twofigw]{mixdist_n4_gu3em6_ang12}
\includegraphics[width=\twofigw]{mixdist_n4_gu3em6_ang24}
\includegraphics[width=\twofigw]{mixdist_n4_gu3em6_ang36}
\includegraphics[width=\twofigw]{mixdist_n4_gu3em6_ang48}
\caption{Distribution of the dipole modulation amplitude $A_\theta$
  obtained for the opening angles $\theta=1.7^\circ$,
  $\theta=6^\circ$, $\theta=12^\circ$, $\theta=24^\circ$,
  $\theta=36^\circ$ and $\theta=48^\circ$ from a mixture of Gaussian
  $\LCDM$ fluctuations and thawing strings with $\Ncorrini=4$
  and $G\U=3\times 10^{-6}$. The pure Gaussian $\LCDM$ distributions
  have been reported for comparison.}
\label{fig:n4_GU3EM6}
\end{center}
\end{figure}

In order to discuss the actual distribution of a mixture of Gaussian
$\LCDM$ anisotropies with delayed scaling strings with $\Ncorrini=4$,
we have chosen a fiducial value of $G\U=3\times 10^{-6}$ (half of the
value taken for $\Ncorrini=2$). Along the same lines as in
section~\ref{sec:mod}, the $1024$ local variance maps of the mixture
have been fitted with a dipole modulation to extract the amplitude
$A_\theta$. The obtained distributions of $A_\theta$ are represented
in figure~\ref{fig:n4_GU3EM6}. As for $\Ncorrini=2$, the distributions
remain very close to the $\LCDM$ ones for angles $\theta \le 2^\circ$
whereas they deviate significantly at larger opening angles. Comparing
figures~\ref{fig:n2_GU6EM6} and \ref{fig:n4_GU3EM6}, one notices that,
in spite of the reduced value of $G\U$ used here, there are more
outliers at large values of $A_\theta$ around intermediate
angles. This is particularly visible in the distribution of $A_6$
($\theta=6^\circ$) and $A_{12}$ ($\theta=12^\circ$). In fact, all of
the eleven local variance maps of figure~\ref{fig:n4_GU3EM6} exhibiting
$A_6 > 0.02$ contain a thawing loop producing one cold spot (four of
them corresponds to the string configurations represented in
figure~\ref{fig:stgmaps4}). On the contrary, for angles $\theta \ge
24^\circ$, the distributions of $A_\theta$ for $\Ncorrini=4$ have a
smaller tail towards large values than for $\Ncorrini=2$. This is the
expected behaviour due to both the smaller value of $G\U$ and the
larger number of strings on the past light cone. The latter effect is
indeed rendering less probable the generation of an anisotropy looking
like a pure dipole on the largest scales as the anisotropy patterns
associated with the higher number of strings start to interfere. The
boosting of the dipole modulation when a loop is present comes from
the averaging effect associated with the local variance map making
procedure, as illustrated in figure~\ref{fig:varmaps2}. The thawing
loop, and its associated cold spot, induces a maximal effect when the
averaging angle captures the typical angular size of the loop. For the
situation discussed here, i.e., $\Ncorrini=4$, this typically occurs
for $\theta$ in the range $6^\circ$-$12^\circ$. Increasing $\Ncorrini$
would decrease the angles at which such an effect takes place.

\section{Conclusion}
\label{sec:conc}

In this paper we have discussed the observational consequences of the
delayed scaling string scenario in which cosmic strings, and possibly
fundamental strings, are generated during inflation. If these objects
are not diluted too much, namely, if they are produced in the last
$60$ e-folds of inflation, they may re-enter the Hubble radius in a
recent past. Using Nambu-Goto numerical simulations in FLRW spacetime,
we have studied the ``thawing regime'' in which strings crossing our
past light cone decouple from the Hubble flow after
recombination. Because visible strings are yet quasi-static and very
rare, they can only distort the CMB in one or two directions in the
sky while the generated patterns may be significantly different than
line discontinuities. Constraints from the angular power spectrum and
direct searches of temperature discontinuities typically allows values
of the thawing string tension to be as large as a few times $10^{-6}$.

By running a thousand of Nambu-Goto cosmic string simulations, we have
generated as many realizations of the CMB sky. We have shown that the
thawing string induced signal mixed with pure Gaussian $\LCDM$
anisotropies can mimic a large scale dipole modulation in the local
variance maps. Both the string configurations having
$\xi/\eta|_\ulss=32$ and $\xi/\eta|_\ulss=16$ can produce a dipole
modulation of amplitude compatible with the observed value $A_\uobs
\simeq 0.06$ provided their tension $G\U =\order{1} \times
10^{-6}$. In addition, from the network configurations having
$\xi/\eta|_\ulss=16$, various realizations exhibit a shrinking loop on
our past light cone whose generic property is to induce a cold spot in
the CMB sky. In the local variance maps, this loop boosts the dipole
modulation amplitude around angular scales close to the actual angular
size of the loop. We have not discussed much smaller correlation
length as we expect these situations to more closely follow the
delayed scaling discussed in Ref.~\cite{Kamada:2014qta}. Indeed,
reducing significantly more the correlation length $\xi$ at last
scattering is expected to produce more loops of smaller angular sizes
while increasing the overall number of visible strings thereby pushing
the statistics closer to the Gaussian situation. At the same time, the
amplitude of the angular power spectrum increases and the
corresponding constraints on $G\U$ would certainly prevent these
strings to be visible at large angles. We have also not discussed the
opposite situation in which $\xi/\eta|_\ulss \gg 32$ because, already
for $\xi/\eta|_\ulss \simeq 32$, around a quarter of all realizations
contain no string intercepting our past light cone. Therefore, larger
correlation lengths at last scattering would correspond to a situation
in which strings remain essentially invisible today. Finally, we left
for a future work any attempt to extract the Bayesian evidence for a
mixture of strings and $\LCDM$ fluctuations to explain both the large
scale dipole modulation and the cold spot. It is indeed a non-trivial
problem as determining any prior distribution, as for instance on the
angular size and number of thawing loops, should rely on a significant
larger number of Nambu-Goto string simulations.

Other discrimination tests could however be envisaged to examine
further if thawing strings are the actual source of the observed large
scale anomalies and the cold spot. For instance, because strings
induce a genuinely non-Gaussian signal, local variance maps should
correlate in a particular way with local skewness or local kurtosis
maps and this could be used to disambiguate thawing strings from other
sources. Moreover, the hypothesis that the cold spot is a shrinking
loop may be tested by examining the lensing patterns of the CMB on its
edge~\cite{Bernardeau:2000xu}. This could already be done by using for
instance small scales CMB telescopes~\cite{Das:2013zf,
  Story:2014hni}. Depending on its shape and velocity, a shrinking
cosmic string loop is expected to induce a warmer CMB temperature in
the adjacent regions located just outside the loop. Another possible
direct detection could be achieved via gravitational lensing
observations by measuring the spatial pattern on the deformation of
photon path, which provides an evidence for the intervening matter
distribution along the line of sight. All photons intercepting the
loop should in addition be redshifted inside and in the same manner
as the CMB. Finally, if the loop possesses kinks, or have enough
dynamics to develop cusps, powerful bursts of gravitational waves
could also be used as tracers~\cite{Damour:2001bk, Olmez:2010bi}.

\acknowledgments The work is supported in part by the
Wallonia-Brussels Federation grant ARC 11/15-040, the Belgian Federal
Office for Science, Technical and Cultural Affairs (C.R.), the
Grand-in-Aid for Japan Society for the Promotion of Science (JSPS)
Fellows No.~259800 (D.Y), and the JSPS Grand-in-Aid for Scientific
Research No.~15H02082 (J.Y).

\bibliographystyle{JHEP}

\bibliography{strings,cmb}

\end{document}